\documentclass{sigplanconf}

\usepackage{amsmath}
\usepackage{amssymb}
\usepackage{stmaryrd}
\usepackage{booktabs}
\usepackage{xspace}
\usepackage{placeins}
\usepackage{enumerate}
\usepackage{subfigure}
\usepackage{verbatim}
\usepackage{enumerate}
\usepackage{algorithm}
\usepackage{algorithmicx}
\usepackage[noend]{algpseudocode}
\usepackage{url}

\usepackage{thm-restate}

\usepackage{pgf, pgfcore, tikz}
\usetikzlibrary{arrows} 
\usetikzlibrary{automata,positioning}
\usetikzlibrary{decorations.pathmorphing}
\usetikzlibrary{shapes.geometric,calc}

\usepackage{multibib}
\newcites{appendix}{References}

\newcommand{\cL}{{\cal L}}

\newcommand{\vpa}{\ensuremath{\mathsf{VPA}}\xspace}

\newcommand{\dvpa}{\ensuremath{\mathsf{DVPA}}\xspace}
\newcommand{\twovpa}{\ensuremath{\mathsf{2VPA}}\xspace}
\newcommand{\dtwovpa}{\ensuremath{\mathsf{D2VPA}}\xspace}
\newcommand{\fcns}{\ensuremath{\mathsf{fcns}}\xspace}
\newcommand{\twovpala}{\ensuremath{\mathsf{2VPA}^{\mathsf{LA}}}\xspace}
\newcommand{\out}{\ensuremath{\mathsf{O}}}
\newcommand{\mout}{\ensuremath{\mathsf{out}}}
\newcommand{\id}{\ensuremath{\mathsf{id}}}

\newcommand{\sem}[1]{\ensuremath{\lbrack \!\lbrack #1 \rbrack \!\rbrack}}

\newcommand{\vpt}{\ensuremath{\mathsf{VPT}}\xspace}

\newcommand{\twovpt}{\ensuremath{\mathsf{2VPT}}\xspace}
\newcommand{\dtwovpt}{\ensuremath{\mathsf{D2VPT}}\xspace}
\newcommand{\twovptla}{\ensuremath{\mathsf{2VPT}^{\mathsf{LA}}}\xspace}
\newcommand{\dtwovptla}{\ensuremath{\mathsf{D2VPT}^{\mathsf{LA}}}\xspace}
\newcommand{\dtwovptlasu}{\ensuremath{\mathsf{D2VPT}_{\mathsf{su}}^{\mathsf{LA}}}\xspace}
\newcommand{\dtwovptsu}{\ensuremath{\mathsf{D2VPT}_{\mathsf{su}}}\xspace}

\newcommand{\VPT}{\vpt}

\newcommand{\biDVPT}{\dtwovpt}
\newcommand{\biDVPTSU}{\dtwovptsu}

\newcommand{\Ha}{\ensuremath{\mathsf{HA}}\xspace}

\newcommand{\dvpt}{\ensuremath{\mathsf{DVPT}}\xspace}

\newcommand{\dHtoS}{\ensuremath{\mathsf{dH2S}}\xspace}
\newcommand{\dHtoSla}{\ensuremath{\mathsf{dH2S}^{\mathsf{LA}}}\xspace}

\newcommand{\MSO}{\ensuremath{\mathsf{MSO}}\xspace}
\newcommand{\MSONW}{\ensuremath{\mathsf{MSO_{nw}}}\xspace}

\newcommand{\MSONWtoW}{\ensuremath{\mathsf{MSO[nw2w]}}\xspace}
\newcommand{\MSOTtoW}{\ensuremath{\mathsf{MSO[b2w]}}\xspace}
\newcommand{\MSOUtoW}{\ensuremath{\mathsf{MSO[u2w]}}\xspace}

\newcommand{\upd}{\mathcal{U}}

\newcommand{\exptime}{{\sc ExpTime}\xspace}

\newcommand{\exptimecomplete}{{\sc ExpTime-c}\xspace}

\newcommand{\ignore}[1]{}

\newcommand{\dom}{\textit{Dom}} 
\newcommand{\pos}{\textit{pos}} 

\newcommand\inter[1]{\llbracket #1 \rrbracket}

\newtheorem{proposition}{Proposition}
\newtheorem{definition}{Definition}
\newtheorem{corollary}{Corollary}

\newenvironment{proof}{\textit{Proof.}}{}
\newenvironment{Sproof}{\textit{Sketch of proof.}}{}
\newtheorem{lemma}{Lemma}
\newtheorem{theorem}{Theorem}
\newtheorem{example}{Example}

\newcommand{\qed}{\hfill\ensuremath{\square}}

\newcommand{\lmark}{\triangleright}
\newcommand{\rmark}{\triangleleft}
\newcommand{\Moves}{\mathbb{D}}
\newcommand{\lmove}{\leftarrow}
\newcommand{\rmove}{\rightarrow}
\newcommand{\move}{\operatorname{move}}
\newcommand{\cread}{\operatorname{read}}
\newcommand{\WN}[1]{\mathcal{N}(#1)}
\newcommand{\stst}{\ensuremath{\mathsf{STST}}\xspace}

\newcommand{\eqclass}[2]{\lbrack #1 \rbrack _{#2}}

\newcommand{\la}{\lambda}

\newcommand{\algwn}{\ensuremath{\mathbb{W}}}
\newcommand{\algtra}[1]{\ensuremath{\mathbb{T}_{#1}}}
\newcommand{\algtrans}[1]{\ensuremath{\mathbb{N}_{#1}}}

\renewcommand{\ll}{\ensuremath{\mathsf{ll}}}
\newcommand{\lr}{\ensuremath{\mathsf{lr}}}
\newcommand{\rl}{\ensuremath{\mathsf{rl}}}
\newcommand{\rr}{\ensuremath{\mathsf{rr}}}

\begin{document}

\setlength{\pdfpageheight}{\paperheight}
\setlength{\pdfpagewidth}{\paperwidth}

\CopyrightYear{2016}
\publicationrights{licensed}
\conferenceinfo{LICS '16,}{July 05 - 08, 2016, New York, NY, USA}
\copyrightdata{978-1-4503-4391-6/16/07}
\reprintprice{\$15.00}
\copyrightdoi{http://dx.doi.org/10.1145/2933575.2935315}


\preprintfooter{2-way Visibly Pushdown Automata \& Transducers}   

\title{Two-Way Visibly Pushdown Automata and Transducers \titlenote{Emmanuel Filiot is research associate at
    FNRS. This work is supported by the ARC project \emph{Transform}
    (French speaking community of Belgium), the Belgian FNRS PDR project
    \emph{Flare}, and the French ANR project ExStream.  
This work has been carried out thanks to the support of the ARCHIMEDE Labex (ANR-11-LABX-0033), the A*MIDEX project (ANR-11-IDEX-0001-02) funded by the Investissements d'Avenir French Government program, managed by the French National Research Agency (ANR) and the PHC project VAST (35961QJ) funded by Campus France and WBI.}
}

\authorinfo{Luc Dartois \and Emmanuel Filiot}
           {Universit\'e Libre de Bruxelles, Belgium}
           {ldartois@ulb.ac.be, efiliot@ulb.ac.be}

\authorinfo{Pierre-Alain Reynier\and Jean-Marc Talbot}
           {Aix-Marseille Universit\'e, CNRS, LIF UMR 7279, 13000, Marseille, France}
           {pierre-alain.reynier@lif.univ-mrs.fr\\ jean-marc.talbot@lif.univ-mrs.fr}
\maketitle

\begin{abstract}


Automata-logic connections are pillars of the theory of regular languages.
Such connections are harder to obtain for transducers, 
but important results have been obtained recently for word-to-word
transformations, showing that the three following
models are equivalent: deterministic two-way transducers, 
monadic second-order (MSO) transducers, and deterministic
one-way automata equipped with a finite number of registers.
Nested words are words with a nesting structure, allowing 
to model unranked trees as their depth-first-search linearisations.
In this paper, we consider transformations from nested words to 
words, allowing in particular to produce unranked trees
if output words have a nesting structure.
%
%
The model of visibly pushdown transducers allows to describe
such transformations, and
we propose a simple deterministic extension of this model with two-way
moves that has the following properties: 
\emph{i)} it is a \emph{simple computational model}, that naturally has a good evaluation complexity;
\emph{ii)} it is \emph{expressive}: it subsumes nested word-to-word MSO transducers, and the 
exact expressiveness of MSO transducers is recovered using a simple syntactic
 restriction;
\emph{iii)} it has \emph{good algorithmic/closure properties}:
the model is closed under composition with a unambiguous one-way letter-to-letter
transducer which gives closure under regular look-around, and has a
decidable equivalence problem.

\end{abstract}



\section{Introduction}\label{Section:Introduction}

\paragraph{Pillars of word language theory }
The theory of languages is one of the deepest and richest theory in
computer science, with successful applications such as, computer-aided
verification and synthesis. A major reason for this success is the 
strong connections between models of languages, with quite different
flavours, that are based on two important pillars: \emph{computation and
logic}. Perhaps one of the most famous example is the effective correspondence
for regular languages of finite words between a low-level
computational model, finite state automata, and a high-level declarative
formalism,  monadic second-order logic (MSO). Similar connections
have been obtained for other structures (e.g. infinite words, finite and
infinite trees, nested words) \cite{Tho97handbook,TATA07}. 
In some cases, it has been
even possible to build a
third pillar based on algebra. The class of regular languages for instance is
known to be the class of languages with finite syntactic congruence.

\paragraph{The logic/two-way/one-way trinity of word transductions} 
To model functions from (input) words to (output) words, i.e. \emph{word
  transductions}, and
more generally word binary relations, automata  have been extended to
\emph{transducers}, i.e. automata with outputs.  Whenever a transducer
reads an input symbol, it can produce on the output a finite word,
the final output word being the right concatenation of all the finite words
produced along the way. To capture functions mirroring or copying
twice the input word, transducers need to read the input word in both
directions: this yields the class of \emph{two-way finite state
  transducers} (2FST).  Two-way transducers have appealing properties:
they are closed under composition \cite{ICALP::ChytilJ1977} and if
they are deterministic, their equivalence problem is decidable (in
\textsf{PSpace}) \cite{Gurari82,CulKar87} and the transduction can be
evaluated in constant space (for a fixed transducer), the output being
produced on-the-fly.  


Impressively, in the late 90s, deterministic two-way
transducers have been shown in
\cite{EngHoo01} to correspond to monadic second-order transducers (MSOT), a
powerful logical formalism introduced in~\cite{Cour94} in a more general
context, with independent motivations. It was the first logic-transducer connection
obtained for a class of transductions with high and desirable expressiveness.
This correspondence
has been extended to finite tree transductions
\cite{MTT,DBLP:journals/siamcomp/EngelfrietM03,DBLP:journals/jcss/BloemE00}.

Recently, an MSOT-expressive one-way model,
\emph{streaming string transducers} (SST), has been introduced in~\cite{alur_et_al:LIPIcs:2010:2853,conf/popl/AlurC11}: it uses registers that can store output words
and can be combined and updated along the run in a linear (copyless)
manner 
The main advantage of
this model is its one-wayness, but the price to pay is the space complexity
of evaluation: it depends also on the size of the register contents.


The models MSOT, deterministic 2FST and deterministic SST have the
same expressive power, and we refer to this correspondence as
\emph{the logic/two-way/one-way trinity}. This trinity has been
extended to transductions of infinite words \cite{DBLP:conf/lics/AlurFT12} and ranked trees
\cite{journals/eatcs/CourcelleE12,DBLP:conf/icalp/AlurD12}.
For trees, bi-directionality is replaced by a tree
walking ability: the transducer can move along the edges of the tree
in any direction. However, to capture MSOT, the transducer needs to
have \emph{regular look-around}, i.e. needs to be able to test regular
properties of the context of the tree node in which it is currently
positioned \cite{journals/eatcs/CourcelleE12}. Look-arounds can be removed at the price of
adding a pushdown store \cite{journals/eatcs/CourcelleE12}. For
one-way machines, uni-directionality is modeled by fixing the
traversal of the tree to be a depth-first left-to-right
traversal and, as for words, to capture MSOT, the transducer needs to
have registers \cite{DBLP:conf/icalp/AlurD12}. Tree-walking transducers with
look-around, and tree transducers with registers are strictly more
expressive than MSOT, but restrictions have been defined that capture
exactly MSOT. Finally, let us mention the macro tree transducers, the
first computational model shown to capture, with suitable
restrictions, MSOT ranked tree transductions
\cite{MTT,DBLP:journals/siamcomp/EngelfrietM03,DBLP:journals/jcss/BloemE00}. This
model has parallel computations, like a top-down tree automaton, and
registers.

\paragraph{Nested words} In this paper, we
consider transductions of nested words to words. \emph{Nested
  words} are words with a nesting structure, built over symbols of
two kinds: call and return symbols\footnote{Sometimes, internal
  symbols are also considered but in this paper, to ease the
  presentation, we omit them. This is wlog as an internal symbol $a$
  can be harmlessly replaced by a call symbol $c_a$ followed by a
  return symbol $r_a$.}. In particular, nested
words can model ordered unranked trees, viewed as their depth-first,
left-to-right, linearisation, and in turn are a natural model of
tree-structured documents, such as XML documents. \emph{Visibly
  pushdown automata} (\vpa) have been introduced in~\cite{journals/jacm/AlurM09} as a model of
regularity for languages of nested words. They are pushdown
automata with a constrained stack policy: whenever a call symbol is
read, exactly one symbol is pushed onto the stack, and when reading a
return symbol, exactly one symbol is popped from the stack. 
Therefore, at any point, the height of the stack corresponds
to the nesting level (call depth) of the word. Roughly, \vpa are 
tree automata over linearised trees, and as such they inherit all the good
closure and algorithmic properties of tree automata. 
However, viewing
trees as nested words has raised motivating questions in the context
of tree streams, such as streaming XML validation \cite{conf/sac/PicalausaSZ11,conf/icdt/SegoufinS07}, streaming
XML queries \cite{conf/www/KumarMV07,journals/iandc/GauwinNT11}, as well as streaming XML transformations
\cite{conf/fsttcs/FiliotGRS11} (see also \cite{webnested} for other applications of \vpa). 

By using a matching predicate $M(x,y)$ that holds true if $x$ is a
call symbol, $y$ is a return symbol and is the matching return of
$x$, MSO logic can be extended from words to nested words, and 
it is known to correspond to regular nested word languages \cite{journals/jacm/AlurM09}.

\paragraph{Nested word to word transductions} Besides the motivations
given before for considering nested words instead of unranked trees,
we argue that seeing unranked trees as nested word yields a natural
and simple two-way model for transductions of nested words, presented later. 
On the output, we do not require the
words to have a particular structure. It is not a weakness: nested
words are words, and the model we introduce in
this paper can as well produce output words that are nested.

\vpa have been extended with output, yielding the class of
\emph{visibly pushdown transducers} (VPT, \cite{conf/mfcs/FiliotRRST10}). When reading an
input symbol, VPT can generate a word on the output. VPT have good
algorithmic and closure properties, and are well-suited to a streaming
context \cite{conf/fsttcs/FiliotGRS11}. However, VPT suffer from a low expressive power, as
they are only one-way, without registers. 

Based on MSO for nested words, one can define MSO transducers \emph{\`a
  la Courcelle} to define nested word to word transductions. From now
on, we refer to such MSO transducers as MSOT. A one-way model has
already been defined in \cite{DBLP:conf/icalp/AlurD12} that captures exactly MSOT. They
extend \vpa with registers that can store partial output
words. Whenever a call symbol is read, the contents of the registers
are pushed onto the stack and all the registers reset. On
reading return symbols, they can combine the content of the current
registers with the content of the registers stored on the stack, in
a copyless fashion. The space complexity of evaluation for 
such transducers is linear in the length of the input nested
word, and they have decidable equivalence problem.

\paragraph{Objective and two-way visibly pushdown transducers} Our main goal in this paper is to
establish a logic/two-way/one-way trinity for nested word to word
transductions. Since the logic/one-way connection has already been
shown in \cite{DBLP:conf/icalp/AlurD12}, we want in particular to define a \emph{two-way
  computational model} with the following requirements: it must be
\emph{conceptually simple}, at least \emph{as expressive as
  MSOT} and have \emph{decidable equivalence problem}.

To this aim, we introduce \emph{deterministic two-way visibly pushdown
  transducers} (\dtwovpt) and show it meets the later requirements. \dtwovpt
read their input in both directions, and their stack behaviour not
only depends on the type of symbols they read, but also on the reading
mode they are in, either backward or forward. In a forward mode, they
behave just like VPT. On the backward mode, they behave like VPT where
the call and return types are swapped: when reading a return symbol
backward, they push a symbol onto the stack, and when reading a call
symbol backward, they pop a symbol from the stack. They can change
their mode at any moment, and produce words on the output.

\begin{figure}
\begin{center}
\begin{tikzpicture}[scale=0.5]

\node (a) at (2,2.5) {$\lmark$};
\node (b1) at (0.5,1) {$2$};
\node (b2) at (1.5,1) {$2$};
\node (b3) at (2.5,1) {$3$};
\node (b4) at (3.5,1) {$1$};

\node (c1) at (-0.3,-0.5) {$3$};
\node (c2) at (0.5,-0.5) {$1$};
\node (c3) at (1.3,-0.5) {$2$};

\foreach \i in {1,2,3,4}
\draw (a) -- (b\i);

\foreach \i in {1,2,3}
\draw (b1) -- (c\i);

\node (fleche) at (4.5,1) {$\Rightarrow$};
\node (e) at (7,2.5) {$\lmark$};
\node (f1) at (5.5,1) {$1$};
\node (f2) at (6.5,1) {$2$};
\node (f3) at (7.5,1) {$2$};
\node (f4) at (8.5,1) {$3$};

\node (g1) at (5.7,-0.5) {$1$};
\node (g2) at (6.5,-0.5) {$2$};
\node (g3) at (7.3,-0.5) {$3$};

\foreach \i in {1,2,3,4}
\draw (e) -- (f\i);

\foreach \i in {1,2,3}
\draw (f2) -- (g\i);

\end{tikzpicture}

\begin{tikzpicture}[scale=0.53]
\newcounter{i};

\foreach \l in { \lmark,2,3,r,1,r,2,r,r,2,r,3,r,1,r,\rmark}
{
\node (letter\arabic{i}) at (\arabic{i},0) {$\l$};
\stepcounter{i};
}


\draw[dashed] (0.5,0.5) -- (12.5,0.5);
\draw[dashed] (14.5,0.5) -- (15,0.5);
\path[dashed] (15,0.5) edge [bend right=90] (15,1);
\draw[dashed] (0,1) -- (15,1);
\path[dashed] (0,1) edge [bend left=90] (0,1.2);
\draw[dashed] (0,1.2) -- (0.5,1.2);
\draw[dashed] (1.5,1.2) -- (3.5,1.2);
\draw[dashed] (5.5,1.2) -- (8,1.2);
\path[dashed] (8,1.2) edge [bend right=90] (8,1.7);
\draw[dashed] (1,1.7) -- (8,1.7);
\path[dashed] (1,1.7) edge [bend left=90] (1,1.9);
\draw[dashed] (1,1.9) -- (5.5,1.9);
\draw[dashed] (8,1.9) -- (7.5,1.9);
\path[dashed] (8,1.9) edge [bend right=90] (8,2.4);
\draw[dashed] (1,2.4) -- (8,2.4);
\path[dashed] (1,2.4) edge [bend left=90] (1,2.6);
\draw[dashed] (1,2.6) -- (1.5,2.6);
\draw[dashed] (3.5,2.6) -- (7.5,2.6);
\draw[dashed] (10.5,2.6) -- (15,2.6);
\path[dashed] (15,2.6) edge [bend right=90] (15,3.1);
\draw[dashed] (0,3.1) -- (15,3.1);
\path[dashed] (0,3.1) edge [bend left=90] (0,3.3);
\draw[dashed] (0,3.3) -- (10.5,3.3);
\draw[dashed] (12.5,3.3) -- (14.5,3.3);
\draw[->] (14.5,3.3) -- (15.5,3.3);

\node (p1) at (0,0.7) {$\lmark$};
\node (p2) at (13,0.7) {$1$};
\node (p3) at (14,0.7) {$r$};
\node (p4) at (1,1.4) {$2$};
\node (p5) at (4,1.4) {$1$};
\node (p6) at (5,1.4) {$r$};
\node (p7) at (6,2.1) {$2$};
\node (p8) at (7,2.1) {$r$};
\node (p15) at (2,2.8) {$3$};
\node (p16) at (3,2.8) {$r$};
\node (p9) at (8,2.8) {$r$};
\node (p10) at (9,2.8) {$2$};
\node (p11) at (10,2.8) {$r$};
\node (p12) at (11,3.5) {$3$};
\node (p13) at (12,3.5) {$r$};
\node (p14) at (15,3.5) {$\rmark$};

\foreach \i/\j in {1/8,2/3,4/5,6/7,9/10,11/12,13/14}
\path (letter\i) edge [bend right=20] (letter\j); 

\foreach \i in {1,2,3,4,5,6,7,8,9,10,11,12,13,15,16}
{
\node (e\i) at ($(p\i) + (0.5,-0.2)$) {};
\draw (p\i)+(-0.5,-0.2) -- (e\i);
}

\end{tikzpicture}

\end{center}
\caption{On top, the transformation of the input. Between siblings
  with the same labeling, the original order is preserved. Below, the
  run of the transducer. Dashed lines are non producing sequences.}\label{Fig:ExRunTri}
\end{figure}
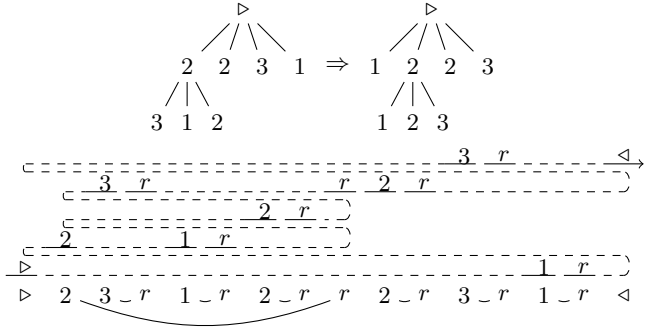

Let us give now an illustrating example of a transduction $f_s$ of nested
words, which will be formalised in Example~\ref{ex:prout}.  
Assume a set of call symbols 
$\{1,\dots,n\}$ ordered by the total order on natural numbers, and one return
symbol $\{r\}$. 
 The transduction $f_s$ sorts an input nested word in
ascending order, recursively nesting level by nesting level, according
to the order on calls. We assume inputs start and end with special symbols
$\lmark$ and $\rmark$ (call and return resp.). E.g., $f_s$ maps $\lmark 22r1rr1r3r \rmark$ to $\lmark
1r21r2rr3r \rmark$ and
$\lmark 23r 1 r 2 r r 2 r 3r 1 r\rmark$ to $\lmark 1 r 2 1 r 2 r 3 r r 2 r 3 r\rmark$ (see Figure~\ref{Fig:ExRunTri}). To make $f_s$ a
function in case the same call symbol occurs twice at the same level,
$f_s$ preserves their order of appearance. The tree
representation of this mapping is given in Figure~\ref{Fig:ExRunTri}
(omitting return symbols). 
The transduction $f_s$ is easily implemented with a \dtwovpt $T_s$. To
process a sequence of siblings at level $k$, $T_s$ works as follows:
for $i$ from $0$ to $n$, $T_s$ performs a forward pass on the siblings
(note that a sibling is actually a tree whose linearisation is of the
form $jwr$ where $w$ is again a sequence of linearised trees). 
During this forward pass, $T_s$ transforms a sibling $jwr$ into
$\epsilon$ if $j\neq i$, and into $iw'r$ otherwise, where $w'$ is the
result of sorting recursively $w$. To implement the loop, when $T_s$
has finished the $i$-th forward pass, i.e. when it reads a return
symbol at level $j-1$, it comes back to its matching call and starts
from there the $(i+1)$-th forward pass, if $i<n$. 

\paragraph{Contributions} By linearising input trees, the simple and
well-known concept of bi-directionality can be generalised naturally
from words to trees. 
While \dtwovpt, as we show in this paper, allow one to lift known results
from word transductions to nested word to word transductions, we
think that \dtwovpt are an appealing model for the following reasons:
\begin{description}
\item[memory efficiency] Regarding the complexity of evaluation,
  for a fixed \dtwovpt, computing the output word of an input nested
  word $w$ can be done in space $O(d(w))$, where $d(w)$ is the depth of
  $w$. Indeed, only the stack and current state need to be kept in
  memory when processing an input nested word. It is an appealing property when transforming large but not
  deep tree-structured
  documents, such as XML documents in general.

\item[expressiveness] At the same time, we show that this efficiency
does not entail expressive power: \dtwovpt can express all MSOT
transductions. They are strictly more expressive than MSOT as they can
for instance express transduction of \emph{exponential size increase},
while MSOT are only of linear size increase. By putting a simple
decidable restriction on \dtwovpt, called \emph{single-useness}, \dtwovpt
capture exactly MSOT transductions. 
    
\item[algorithmic properties] 
Despite their high expressive power,
  \dtwovpt still have \emph{decidable equivalence problem}. 
We also prove that preprocessing the input of a \dtwovpt by a letter-to-letter unambiguous \vpt does not increase its expressive power, as their composition is again a \dtwovpt.
\end{description}

The proof of expressiveness relies on an existing correspondence
between tree-walking and MSO transducers of ranked trees to
words~\cite{journals/eatcs/CourcelleE12}, and on the classical
\emph{first child-next sibling} (fcns for short) encoding of unranked
trees into binary trees.  As in~\cite{journals/eatcs/CourcelleE12}, we
use an intermediate automata model equipped with MSO look-around, and then show that
these look-around tests can be removed. For the latter property, our proof
differs from that of~\cite{journals/eatcs/CourcelleE12} in which 
a pushdown stack is used to update information on MSO-types.  
On binary trees, their model pushes the stack while moving to the
first-child, but also while moving to the second child. This latter push corresponds,
through the fcns encoding, to pushing a symbol while moving to the next
sibling, an operation that is not allowed with a visibly pushdown stack. 
Hence, in order to prove that look-around tests can be removed in our model, 
we need a more involved construction, that extends a non-trivial result 
proven in~\cite{HopUll67} for two-way automata on words. %
%
Decidability of \dtwovpt equivalence is done by reduction to
deterministic top-down tree to word transducer equivalence, a problem
which was opened for long and recently solved in~\cite{SMK-FOCS15}.

\paragraph{Application 1: Unranked tree to word walking transducers}%
\hspace{-3mm}\dtwovpt can easily be translated into a pushdown walking model of
unranked tree to word transductions.  It works exactly as in the
ranked tree case of~\cite{journals/eatcs/CourcelleE12}: one stack
symbol is pushed while going downward and popped while going
upward. While moving along sibling relations, the stack is
untouched. As a consequence of our results, this model, with
single-use restriction, captures exactly MSOT. This model is discussed in the last section.

\paragraph{Application 2: Query 2VPA}
 Deterministic two-way \vpa have been introduced in
 \cite{DBLP:conf/mfcs/MadhusudanV09} as an equi-expressive 
 model for MSO-definable
 unary queries on nested words. Using
 \cite{NevSch02,conf/dbpl/NiehrenPTT05}, such queries can be shown to be equivalent to
 unambiguous \vpa with special states which select the nested word positions that
 are answers to the query. As shown in
 \cite{DBLP:conf/mfcs/MadhusudanV09},  unambiguity can be
 traded for determinism, at the price of adding two-wayness. This result comes as
 a consequence of ours: a one-way unambiguous selecting \vpa can be
 seen as a deterministic \vpt with look-around, that annotates the
 input positions selected by the \vpa (look-around resolves
 nondeterminism), which can be transformed into a \dtwovpt using our
 results. The main ingredient of the proof
 of~\cite{DBLP:conf/mfcs/MadhusudanV09} is also a Hopcroft-Ullman
 construction, but in a setting simpler than ours~\footnote{They
   provide a construction for the composition of a co-deterministic \vpa with an unambiguous \vpa, 
while we study that of a \dtwovpa with an unambiguous \vpa.}.

\paragraph{Organisation of the paper}
In Section \ref{Section:Twovpa}, we introduce two-way \vpa and two-way \vpa with
look-around, define the notion of transition algebra for \twovpa and use this to show that they are equivalent to one-way \vpa. As a
consequence, they have decidable (exptime-c) emptiness problem.
In Section \ref{Section:Twovpt}, we introduce \dtwovpt and \dtwovpt with look-around, show
that they are equivalent, and study their algorithmic
properties. Section \ref{Section:Expressiveness} is devoted to the expressiveness of \dtwovpt,
with a comparison to MSOT and to other known models of nested word to
word transductions. Due to lack of space, some results are proved in
Appendix. Finally, all our expressiveness equivalences are effective.


\section{Two-way visibly pushdown automata}\label{Section:Twovpa}

\subsection{Definitions}

We introduce in this section two-way visibly pushdown
automata, following the definition of \cite{DBLP:conf/mfcs/MadhusudanV09}.

We consider a structured alphabet $\Sigma$ defined as the disjoint
union of call symbols $\Sigma_c$ and return symbols
$\Sigma_r$.  The
set of words over $\Sigma$ is $\Sigma^*$. As usual, $\epsilon$
denotes the empty word. Amongst words, the set of nested
words $\WN{\Sigma}$ is defined as the least set such that $\epsilon
\in \WN{\Sigma}$ and if $w_1,w_2 \in \WN{\Sigma}$ then both $w_1w_2$ 
and $cw_1r$ (for all $c \in \Sigma_c$ and $r \in \Sigma_r$) belong to
$\WN{\Sigma}$. 
In the following, we assume that input words of our models are always nested words.
This is not restrictive as all our models can recognize and filter nested words.

For a word $w\in\Sigma^*$, its length is denoted by $|w|$ and we denote by $w(i)$
its $i$th symbol. Its set of positions is $\pos(w) = \{1,\dots,|w|\}$,
and for $i,j\in \pos(w)$ such that $i<j$,  we say that $(i,j)$ is a
\emph{matching pair} of $w$ if $w(i)\in \Sigma_c$, $w(j)\in\Sigma_r$ and 
$w$ can be decomposed into $w= w_1 w(i) w_2 w(j) w_3$, where $w_1,w_3\in\Sigma^*, w_2\in
\WN{\Sigma}$ and $|w_1| = i-1$, $|w_2| = j-i-1$. Note that if
$w\in\WN{\Sigma}$, then necessarily, $w_1w_3\in\WN{\Sigma}$.

When dealing with two-way machines, we assume the structured alphabet $\Sigma$
to be extended into $\overline{\Sigma}$ by adding two special symbols
$\lmark,\rmark$ in $\overline{\Sigma}_c$ and $\overline{\Sigma}_r$
respectively, and we consider words with left and right markers from
$\lmark \Sigma^* \rmark$.

\begin{definition}
  A \emph{two way visibly pushdown automaton}
  (\twovpa for short) $A$ over $\overline{\Sigma}$ is given by
  $(Q,q_I,F,\Gamma,\delta)$ where $Q$ is a finite set of states, $q_I
  \in Q$ is the initial state, $F \subseteq Q$ is a set of final states and
  $\Gamma$ is a finite stack alphabet. Given the set $\Moves=\{\lmove,
  \rmove\}$ of directions, the transition relation $\delta$ is defined by 
$\delta^{\text{push}} \cup \delta^{\text{pop}}$ where 
\begin{itemize}
\item  $\delta^{\text{push}} \subseteq ((Q \times \{\rmove\} \times \Sigma_c)  \cup 
(Q \times \{\lmove\} \times \Sigma_r)) \times ((Q \times \Moves) \times \Gamma)$
\item $\delta^{\text{pop}} \subseteq  ((Q \times \{\lmove\} \times \Sigma_c
  \times \Gamma) 
  \cup  (Q \times \{\rmove\} \times \Sigma_r \times \Gamma)) \times (Q \times \Moves)$
\end{itemize}
Additionally, we require that for any states $q,q'$ and any stack symbol $\gamma$, if
$(q,\lmove,\lmark,\gamma,q',d) \in \delta^{\text{pop}}$ then $d=\rmove$ and if 
$(q,\rmove,\rmark,\gamma,q',d) \in \delta^{\text{pop}}$ then $d=\lmove$.
\end{definition}

Informally, a \twovpa has a reading head pointing between symbols (and
possibly on the left of $\lmark$ and on the right of $\rmark$). A
configuration of the machine is given by a state, a direction $d$ and
a stack content. The next symbol to be read is on the right of the
head if $d=\rmove$ and on the left if $d=\lmove$.  Note that when
reading the left marker from right to left $\lmove$ (resp. the right
marker from left to right $\rmove$), the next direction can only be
$\rmove$ (resp. $\lmove$). The structure of the alphabet induces the
behaviour of the machine regarding the stack when reading the input
word: when reading on the right, a call symbol leads to push onto the
stack while a return symbol pops a symbol from the stack. When reading
on the left, a dual behaviour holds (hence, at a given position in
the input word, the height of the stack is always constant at
each visit to that position in the run). Finally, the state and the
direction are updated. 

Let $w\in\WN{\Sigma}$. We set $w(0) = \lmark$ and $w(|w|+1) =
\rmark$. For a move $d$ and $0\leq i \leq |w|$, we denote by
\begin{itemize}
\item $\move(d,i)$ the integer $i-1$ if $d=\lmove$ and $i+1$ if
  $d=\rmove$.
\item $\cread(w,d,i)$ the symbol $w(i)$ if $d=\lmove$ and $w(i+1)$ if 
$d=\rmove$.
\end{itemize}

Formally, a stack $\sigma$ is a finite word over $\Gamma$. The empty
stack/word over $\Gamma$ is denoted $\bot$. For a word $\lmark
w\rmark$ where $w\in\WN{\Sigma}$ and a \twovpa $A=(Q,q_I,F,\Gamma,\delta)$, a
\emph{configuration} of $A$ is a triple $(q,i,d,\sigma)$ where $q \in
Q$, $0 \le i \le |w|+1$, $d \in \Moves$ and $\sigma$ is a stack. A
\emph{run} of $A$ on a word $w$ is a finite non-empty sequence of
configurations $(q_0,i_0,d_0,\sigma_0) (q_1,i_1,d_1,\sigma_1) \ldots
(q_\ell,i_\ell,d_\ell,\sigma_\ell)$ where for all $0\le j \le \ell$,
the configuration $(q_{j+1},i_{j+1},d_{j+1},\sigma_{j+1})$ satisfies
$i_{j+1}=\move(i_j,d_j)$ and
\begin{itemize}
\item if $\cread(w,d_j,i_j) \in \Sigma_c$ and $d_j=\rmove$  or 
$\cread(w,d_j,i_j) \in \Sigma_r$ and $d_j=\lmove$ then 
$(q_j,d_j,\cread(w,d_j,i_j),q_{j+1},d_{j+1},\gamma) \in \delta^{\text{push}}$
and $\sigma_{j+1}=\sigma_j\gamma$. 
\item  if $\cread(w,d_j,i_j) \in \Sigma_c$ and $d_j=\lmove$  or 
$\cread(w,d_j,i_j) \in \Sigma_r$ and $d_j=\rmove$ then 
$(q_j,d_j,\cread(w,d_j,i_j),\gamma,q_{j+1},d_{j+1}) \in \delta^{\text{pop}}$
and $\sigma_{j+1}\gamma=\sigma_j$. 
\end{itemize}
By the special treatment of $\lmark$ and $\rmark$ ensured by the
definition of \twovpa, the indices $i_j$ all belong to $\{0,\dots,|w|+1\}$.
Note also that any configuration is actually a run on the empty word
$\epsilon$.
A run on a nested word $w$ is accepting whenever  $q_0= q_I$, $i_0=0$, 
$d_0=\rmove$, $\sigma_0=\bot$ and $q_\ell \in F$, $i_\ell=|w|+1$,
$d_\ell=\rmove$, $\sigma_\ell=\bot$.  

Note that $A$ being a visibly pushdown automaton, for any two
configurations in a run of $A$ at the same position $i$ in the word
$(q,i,d,\sigma)$ and $(q',i,d',\sigma')$, the stack $\sigma$,
$\sigma'$ have the same height/length. 

The language $\cL(A)$ defined by $A$ is the set of
nested words $w$ from $\Sigma^*$ such that
there exists an accepting run of $A$ on
 $\lmark w \rmark$.

\begin{definition}
A two-way visibly pushdown automaton is 
\begin{itemize}
\item \emph{deterministic} (\dtwovpa for short)
if we may write $\delta^{\text{push}}$,
$\delta^{\text{pop}}$ as functions from $((Q \times \{\rmove\} \times
\Sigma_c) \cup (Q \times \{\lmove\} \times \Sigma_r))$ to $(Q \times
\Moves) \times \Gamma$ and from $((Q \times \{\lmove\} \times \Sigma_c
\times \Gamma) \cup (Q \times \{\rmove\} \times \Sigma_r \times
\Gamma))$ to $Q \times \Moves$ respectively.
%
\item \emph{codeterministic} if we may write $\delta^{\text{push}}$,
$\delta^{\text{pop}}$ as injective applications, with the same type as in the previous item.

\item \emph{unambiguous} iff for any word $w$, there exists at most one
accepting run on $w$. 
\end{itemize}
\end{definition}

Obviously, if $A$ is (co)deterministic, for any word $w$ from
$\WN{\bar{\Sigma}}$, there exists a unique run on $w$ in $A$ from any
fixed configuration. Hence, any (co)deterministic \twovpa is
unambiguous. Note also that the determinism of $A$ implies that any
configuration can occur only once in some accepting run (otherwise,
the machine would loop without reaching a final configuration).
 
A two-way visibly pushdown automaton is a \emph{(one-way) visibly pushdown
automaton} (\vpa for short) whenever $d'=d=\rmove$ for all
$(q,d,\alpha,q',d',\gamma')$ in $\delta^{\text{push}}$ and for all 
$(q,d,\alpha,\gamma,q',d')$ in $\delta^{\text{pop}}$.

For \vpa, we  may omit directions in the transition relation, configurations and runs. 

Finally, we will denote $\dvpa$ the class of deterministic \vpa. In
this case, the transition relation is defined as a function omitting
directions.

\subsection{Transition algebra for \twovpa} 

Nested words from $\WN{\Sigma}$ (or $\WN{\overline{\Sigma}}$)
induce a natural algebra $\algwn=(\WN{\Sigma},.,\{f_{c,r} \mid c \in
\Sigma_c,\ r \in \Sigma_r\},\epsilon)$ where '$.$' is a binary
operation, the $f_{c,r}$ form a family of unary operations and
$\epsilon$ is a constant. The semantics of $\epsilon$ is the empty
word, of $.$ is concatenation and for any $w$ in $\WN{\Sigma}$,
$f_{c,r}(w)=cwr$. Obviously, the operators finitely generates
$\WN{\Sigma}$ which can be seen as the free generated algebra over
this signature quotiented by the associativity of '$.$' and the
neutrality of $\epsilon$ wrt the concatenation '$.$'.

\paragraph{The traversal congruence $\sim$}
Inspired by works on two-way automata on words~\cite{Pecuchet85,Shepherdson59},
we study traversals of a \twovpa $A$.
A traversal of some nested word $w$ abstracts a run of $A$ keeping
track only of the fact that it starts reading the word from the left
or from the right (depending on the initial direction) in some state
$p$ and leaves it in some state $q$. Now, formally, for any states
$p,q$, and any two directions $d_1,d_2 \in \Moves$,
$((p,d_1),(q,d_2))$ belongs to the traversal of $w$ if there exists a
run of $A$ on $w$ starting in the configuration $(p,pos(d_1),
d_1,\bot)$ and ending in $(q,pos(d_2), d_2,\bot)$, where
$$
\left \{
\begin{array}{l}
pos(d_1) = 0 \text{~~if~} d_1=\rmove \quad \text{~and~} \quad pos(d_1) = |w| \text{~~otherwise}\\
pos(d_2) =|w| \text{~~if~} d_2=\rmove \quad \text{~and~} \quad pos(d_2) = 0 \text{~~otherwise}
\end{array}
\right .
$$

Note that the reading starts either at the beginning or at the end of
$w$ depending on the initial current direction and that the final
direction indeed leads to leave the word.  One may associate with a
nested word the set of its traversals and define a relation
$\sim$ on nested words such that $u \sim v$ if $u$ and $v$ have
the same traversals.

Obviously, $\sim$ is an equivalence relation over $\WN{\Sigma}$ and we
denote by $\lbrack w\rbrack_{\sim}$ the set of traversals of a
nested word $w$. We prove that $\sim$ is actually a
congruence, that is if $w_1 \sim w_2$ and $w'_1 \sim w'_2$ then
$f_{c,r}(w_1)= cw_1r \sim cw_2r = f_{c,r}(w_2)$ and $w_1.w'_1 \sim
w_2.w'_2$ for any nested words $w_1,w'_1,w_2,w'_2$ in
$\WN{\Sigma}$. 
\begin{restatable}{proposition}{propSimCongruence}
\label{prop:congruencetraversal}
The relation $\sim$ is a congruence of finite index. 
\end{restatable}

\paragraph{The transition algebra $\algtra{A}$}
Based on Proposition \ref{prop:congruencetraversal}, the congruence
relation $\sim$ induces a finite algebra
$\algtra{A}=(\text{Trav}_A,.^{\algtra{A}}, \{f^{\algtra{A}}_{c,r} \mid
c \in \Sigma_c,r \in \Sigma_r\}, \epsilon^{\algtra{A}})$ where the
support is $\text{Trav}_A$ the set of all traversals induced by $A$,
$.^{\algtra{A}}$ is a binary operation which is associative, each
$f^{\algtra{A}}_{c,r}$ is a unary operation and
$\epsilon^{\algtra{A}}$ is a constant from $\text{Trav}_A$ and a
neutral element for $.^{\algtra{A}}$. More specifically,
$\epsilon^{\algtra{A}} = \eqclass{\epsilon}{\sim}$,
$\eqclass{u}{\sim}.^{\algtra{A}}\eqclass{v}{\sim} = \eqclass{uv}{\sim}$ and 
$f^{\algtra{A}}_{c,r}(\eqclass{u}{\sim}) = \eqclass{cur}{\sim}$. These operations
are well-defined since $\sim$ is a congruence. 


Hence, there exists a unique and canonical 
morphism $\mu_{\algtra{A}}$ from $\algwn$, the algebra of nested
words, onto $\algtra{A}$, that satisfies $\mu_{\algtra{A}}(w)=
\eqclass{w}{\sim}$. We also denote $\eqclass{w}{\sim}$ as
$w^{\algtra{A}}$ since it can be considered as the interpretation
of $w$ (which is an element $\algwn$) in $\algtra{A}$. 

The correction of this morphism 
$\mu_{\algtra{A}}$ directly implies:
\begin{proposition}
\label{cor:2vpa2alg}
  Let $A=(Q,q_I,F,\Gamma,\delta)$ be a \twovpa. 
$\mathcal{L}(A)=\mu_{\algtra{A}}^{-1}(\{m \in \text{Trav}_A \mid m \cap
(\{(q_I,\rmove)\} \times F \times \{\rmove\}) \neq \varnothing\})$.
\end{proposition}
%

Note that this statement corresponds to the classical notion of recognizability by some
finite algebra.

\subsection{From two-way visibly pushdown automata to visibly pushdown automata}
In this subsection we give a reduction from \twovpa to \vpa.
While this result can be inferred from~\cite{DBLP:conf/mfcs/MadhusudanV09}, our Shepherdson-inspired approach gives an upper bound on the complexity of the procedure.
We first recall the notion of recognizability by finite algebra and
show that this notion is equivalent to recognazibility by \dvpa. Then
we prove the main result of this section appealing to the transition
algebra $\algtra{A}$.


Let $\mathbb{A}=(D_\mathbb{A},.^\mathbb{A},(f^\mathbb{A}_{c,r})_{(c,r)
\in \Sigma_c \times \Sigma_r},\epsilon^\mathbb{A})$ be a finite
algebra such that $.^\mathbb{A}$ is associative having 
$\epsilon^\mathbb{A}$ as neutral element. 
There exists a unique morphism $\mu_\mathbb{A}$ from the algebra of
nested words $\algwn$ onto $\mathbb{A}$.

\begin{definition}
A language $\cL \subseteq \WN{\Sigma}$ is recognized
by $\mathbb{A}$ if there exists a set $\cL_{\mathbb{A}} \subseteq
D_\mathbb{A}$ such that $\cL=\mu_\mathcal{A}^{-1}(\cL_\mathbb{A})$.  
\end{definition}

As an example, as shown in Proposion \ref{cor:2vpa2alg}, a language $\cL$ defined
by a \twovpa is recognized by the transition algebra $\algtra{A}$. We
show that recognability by finite algebra implies \dvpa
recognizability. 

\begin{lemma}
\label{lem:alg2dvpa}
If $\cL$ is recognized by a finite algebra $\mathbb{A}$ then it is
recognizable by a \dvpa $B_\mathbb{A}$. Moreover, the size of
$B_\mathbb{A}$ is polynomial in the size of $D_\mathbb{A}$,
the support of $\mathbb{A}$.
\end{lemma}

\begin{proof} 
For $\mathbb{A}$ and the set $\cL_\mathbb{A} \subseteq D_\mathbb{A}$, we define the \dvpa
$B_{\mathbb{A}}=(D_\mathbb{A},\epsilon^{\mathbb{A}},\cL_\mathbb{A},\Sigma_c
\times D_\mathbb{A}, 
\delta_{B_\mathbb{A}})$ where  $\delta_{B_\mathbb{A}} = \delta^{\text{push}}_{B_\mathbb{A}} \cup
  \delta^\text{pop}_{B_\mathbb{A}}$ and $\delta^{\text{push}}_{B_\mathcal{A}}(m^{\mathbb{A}},c)=(\epsilon^{\mathbb{A}},(c,m^{\mathbb{A}}))$,
$\delta^{\text{pop}}_{B_\mathcal{A}}(m'^{\mathbb{A}},r,(c,m^{\mathbb{A}}))=m^{\mathbb{A}}
\circ f^{\mathbb{A}}_{c,r}(m'^{\mathbb{A}})$. 
Obviously, $B_\mathbb{A}$ is deterministic. Its correctness can be
proved by induction on nested words showing for all $w \in
\WN{\Sigma}$, there exists a run in $B_\mathbb{A}$ on $w$ from
$(m^\mathbb{A},0,\bot)$ to $(m'^{\mathbb{A}},|w|,\bot)$ iff
${m}'^{\mathbb{A}}=m^{\mathbb{A}} .^{\mathbb{A}} \mu_{\mathbb{A}}(w)$.
And so, for an accepting run on $w$ from $(\epsilon^\mathbb{A},0,\bot)$ to
$(m'^{\mathbb{A}},|w|,\bot)$
with $m'^{\mathbb{A}} \in \cL_\mathbb{A}$,
${m}'^{\mathbb{A}}=\mu_{\mathbb{A}}(w)$. Hence,
$\cL(B_{\mathbb{A}})= \mu_{\mathbb{A}}^{-1}(\cL_{\mathbb{A}})$. 
Finally, note that the number of states of $B_\mathcal{A}$ is precisely the
cardinality of the support of $\mathcal{A}$. \qed
\end{proof}

We can now come to the main result of this section.
\begin{theorem}
\label{thm:2vpa2dvpa}
For any $\twovpa$ $A$, one can compute (in exponential time) a $\dvpa$ $B$ such that 
$\cL(A)=\cL(B)$ and the size of $B$ is exponential in the
size of $A$. 
\end{theorem}
\begin{proof}
  One can build from the $\twovpa$ $A$ the elements of
  $\{\eqclass{w}{\sim} \mid w \in \WN{\Sigma}\}$ and thus, the
  transition algebra $\algtra{A}$, in exponential time.  Then, by
  Lemma \ref{lem:alg2dvpa}, a \vpa $B_{\algtra{A}}$ is built
  from $\algtra{A}$. The correctness follows from Proposition
\ref{cor:2vpa2alg} for $\cL_{\algtra{A}}= \{m^{\algtra{A}} \in \text{Trav}_A \mid m^{\algtra{A}}  \cap
(\{(q_I,\rmove)\} \times F \times \{\rmove\}) \neq \varnothing\}$. \qed 
\end{proof}

\begin{restatable}{corollary}{coremptinesstwovpa}
\label{Cor-Emptiness2VPA}
For any $\twovpa$ $A$, deciding the emptiness of $A$ (\textsl{ie}
$\cL(A)=\varnothing$) is \exptimecomplete. The same result
holds for $\dtwovpa$. 
\end{restatable}
\begin{proof}
We prove the upper-bound for \twovpa and the lower bound for
\dtwovpa. 
For the upper-bound, it suffices to build from $A$ in exponential time 
a  equivalent \dvpa $B$ possibly exponentially larger than $A$ (Theorem
\ref{thm:2vpa2dvpa}). Then, emptiness of $B$ can be tested in polynomial
time \cite{journals/jacm/AlurM09}. 

The proof of the lower bound proceeds by a reduction of the
emptiness problem of intersection of $k$ deterministic
top-down tree automata, that is known to be \exptimecomplete.

\end{proof}

\subsection{\twovpa with look-around}

As we will later on need the notion of look-around for transducers, we
introduce it first for automata to ease the presentation. Hence, we
extend the model of \twovpa with look-around. The feature will add a
guard to each transition of the machine. This guard will require to be
satisfied for the transition to be applied.

\begin{definition}
  A \twovpa with look-around (\twovpala for short) is given by a
  triple $(A,\la,B)$ such that $A$ is a \twovpa and $B$ a unambiguous
  \vpa and $\la$ is a mapping from the transitions of $A$ to the
  states of $B$.
\end{definition} 

The notion of runs is adapted to take into account look-around as
follows: in any run on some nested word $w$, for any two successive configurations 
$(q_j,i_j,d_j,\sigma_j) (q_{j+1},i_{j+1},d_{j+1},\sigma_{j+1})$ obtained by a transition $t$, we require
that there exists a unique accepting run on $w$ in $B$ and that this run contains a
configuration of the form $(\lambda(t),\cread(w,d_j,i_j),\sigma)$. 

The definition of accepting runs remains the same and the language defined by
such machines is defined accordingly.

The notion of one-wayness extends trivially to \twovpa with
look-around. For determinism, we ask the look-around to be disjoint
on transitions with the same left hand-side: for any two different transitions
of $A$, $t_1=(q,d,a,q'_1,d'_1,\gamma_1)$,
$t_2=(q,d,a,q'_2,d'_2,\gamma_2)$ in $\delta_c$ (resp.
$t_1=(q,d,a,\gamma,q'_1,d'_1)$, $t_2=(q,d,a,\gamma,q'_2,d'_2)$
in $\delta_r$), it holds that $\la(t_1)\neq\la(t_2)$.

Non-surprisingly, \twovpa are closed under look-around:

\begin{theorem} 
Given a \twovpala $(A,\la,B)$, there exists a \vpa $A'$ such that 
$\cL((A,\la,B))=\cL(A')$.  
\end{theorem}




\section{Two-way visibly pushdown transducers}\label{Section:Twovpt}

\subsection{Definitions}

Let $\Sigma,\Delta$ be two finite alphabets such that $\Sigma$ is structured.
Two-way visibly pushdown transducers (\twovpt) from $\Sigma$ to $\Delta$ extend 
\twovpa over $\Sigma$ with a one-way-left-to-right output tape.
They are defined
as a pair $T=(A,\out)$ where $A$ is a \twovpa over $\Sigma$ and $\out$ is a
morphism from the set of rules of $A$ to words in $\Delta^*$. 

A run of a \twovpt $T=(A,\out)$ on an input word $w\in \WN{\Sigma}$ is a run $\rho$ of $A$ on $w$.
We say the run is accepting if it is in $A$.
A run $\rho$ may be simultaneously a run on a word $w$ and on a word $w'\neq w$,
however, when the underlying input word $w$ is given, there is a unique sequence of 
transitions $t_1t_2 \ldots t_n$ associated with $\rho$ and $w$. In this case, the 
output produced by the run $\rho$ on $w$ is defined as the word 
$v=\out(t_1)\out(t_2)\ldots \out(t_n)\in\Delta^*$. This word is denoted by $\mout^w(\rho)$.
If $\rho$ contains a single configuration, then we let $\mout^w(\rho)=\epsilon$.
The transduction defined by $T$ is the relation $$\inter{T}=\{(w,\mout^w(\rho)) \in \WN{\Sigma}\times\Delta^* \mid 
\rho \text{ is an accepting run of }T\text{ on }w\}.$$
We say that $T$ is \emph{functional} if $\inter{T}$ is a function, and that $T$ is 
\emph{deterministic} (resp. \emph{unambiguous})
if $A$ is deterministic (resp. unambiguous). The class of deterministic two-way visibly pushdown transducers
is denoted \dtwovpt. Observe that if $T$ is deterministic or unambiguous, then it is trivially
functional. Last, when $T$ is functional, we may interpret the relation $\inter{T}$
as a partial function on $\WN{\Sigma}$: given a word $w\in \WN{\Sigma}$, denote by 
$\inter{T}(w)$ the unique word $v\in \Sigma^*$ such that $(w,v)\in \inter{T}$,
whenever it exists.
To ease readability, we may simply write $T$ to denote $\inter{T}$ when it is clear from the context, for example when considering composition of functions.

We consider classes of one-way visibly pushdown transducers, obtained
by considering the corresponding classes of one-way visibly pushdown automata.
The notions of functional, deterministic and unambiguous transducers are
naturally defined for these transducers, and we denote by (D)\vpt 
the class of (deterministic) one-way visibly pushdown transducers. 
Last, we say that a \vpt $T=(A,\out)$ from $\Sigma$
to $\Delta$ is \emph{letter-to-letter}
if $\Delta$ is a structured alphabet and if $\out$ maps
every call transition of $A$ to an element of $\Delta_c$
and every return transition of $A$ to an element of $\Delta_r$.

\twovpt (resp. \dtwovpt) can be extended with look-around, as we did for \twovpa.
Formally, a two-way visibly pushdown transducer with look-around (\twovptla for short)
is a pair $T=(A',\out)$ where $A'=(A,\lambda,B)$ is a \twovpala and $\out$
is a morphism from the set of rules of $A$ to words in $\Delta^*$.
We say that such a machine is deterministic if the \twovpala $A$ is deterministic, the resulting
class being denoted by \dtwovptla.

\begin{example}\label{ex:prout}
We now formally express the transduction given in the introduction (see Figure~\ref{Fig:ExRunTri}).
Let $Q = \{ q_1,\dots,q_n \} \cup
\{q_{i,j}\mid 1\leq i,j\leq n\text{ or }i=\lmark\}\cup \{q_f\}$ be the set of states  with initial state $q_1$ and final state
$q_f$, a set of stack symbols $\Gamma = \{\bot\}\cup \{i\mid
i=1,\dots, n\}$, and for all $i,j,k\in\{\lmark,1,\dots,n\}$, we have the rules:
%
{\small
$$
\begin{array}{rclrcc}
q_i,\rmove & \xrightarrow{i\mid i, +i} & q_1,\rmove &
 q_n,\rmove & \xrightarrow{r\mid r, -j} & q_j,\rmove\\

  q_i,\rmove & \xrightarrow{(j,r)} & q_i,\rmove\text{ if }
                                                 j\neq i &
 q_{i,j},\lmove & \xrightarrow{(j,r)} & q_{i,j},\rmove \\
q_i,\rmove & \xrightarrow{r\mid \epsilon, -j} & q_{i,j},\lmove\text{ if } i < n
 &
 q_{i,j},\rmove & \xrightarrow{k|\epsilon, +j} & q_{i+1},\rmove
\end{array}
$$}
The markers are treated as letters, except that they push $\bot$ instead of $\lmark$ and
upon popping $\bot$ in state $q_n$, the transducer goes to $q_f$ and accepts.
The transitions labeled by $(j,r)$ are macros corresponding to moves along matching relation, which can easily be implemented.
\end{example}

\paragraph{Evaluation}

Observe here that if a transformation is given as a \dtwovpt $T$, then one can evaluate it using
a memory linear in the depth of the input word $w$ (we assume $w$ can be accessed as we want
on some media). Indeed, one simply needs to store the current configuration of $T$,
given as a state and a stack content.


\subsection{Closure under composition}

We prove in this subsection that $\twovpt$ are closed by composition
with a letter-to-letter unambiguous \vpt, extending a similar result
for transducers on words~\cite{HopUll67}. This will reveal useful to
show that \dtwovpt are closed under
look-around. First, we extend to nested words a result that was known for finite
transducers:
%
\begin{restatable}{lemma}{lemDecompUnVPT}\label{lemma-DecompUnambVPT}
Any unambiguous \vpt $T$ can be written as the composition of two \vpt 
$T_1\circ T_2$, where $T_1$ is deterministic and $T_2$ is letter-to-letter and co-deterministic.
Furthermore, if $T$ is letter-to-letter, so is $T_1$.
\end{restatable}

\begin{restatable}{theorem}{ThmHUVPT}\label{Thm-HUVPT}
Given a letter-to-letter $\dvpt$ $A$ and a $\twovpt$ $B$, we can construct a $\twovpt$ $C$ that realizes the composition $C=B\circ A$.

If furthermore $B$ is deterministic, then so is $C$. 
\end{restatable}

\begin{proof}
We first notice that since we are considering visibly pushdown machines and the first machine is letter-to-letter, the stacks of both machines are always synchronized, meaning that they have the same height on each position.
Then, let us remark that when the $\twovpt$ moves to the right, 
we can do the simulation in a straight forward fashion by simulating it on the production of the one-way.
It becomes more involved when it moves to the left.
We then need to rewind the run of the one-way, and nondeterminism can arise.
To bypass this, let us recall that a similar construction from~\cite{HopUll67} exists for classical transducers, and that the rewinding is done through a back and forth reading of the input, backtracking the run up to a position where the nondeterminism is cleared, and then moving back to the current position.
The method is to compute the set of possible candidates for the
previous state, and keep moving to the left until we reach a position
$i$ where there is only one path left leading to the starting position $j$.
Afterward, we simply follow this path along another one from position
$i+1$. As we know that they will merge at position $j$, we can stop at position $j-1$ with the correct state.
If we reach the beginning of the word with multiple candidates, we do the same procedure, the correct path being the one starting from the initial state.

This cannot be done as such on pushdown transducers since rewinding the run might lead to popping the stack, and losing information.
However, if at each push position, we push not only the stack symbols but also the current state, we are able, when rewinding the run, to clear the nondeterminism as soon as we pop this information by using it as a \textit{local initial state}, limiting the back and forth reading to the current subhedge.
The overall construction can be seen as a classical Hopcroft-Ullman construction on hedges, abstracted as words over the left-to-right traversals of
their subhedges, which are called summaries in
\cite{journals/jacm/AlurM09} (see Figure~\ref{Figure:WordSubhedge}).
These summaries can be computed on-demand by a one-way automaton.

Finally, note that to apply this construction, we need to push this local initial state each time we enter a subhedge, whether we enter from the right or from the left.
This can be maintained since when entering from the left, it simply corresponds to the current state
and when entering from the right, this state is computed by the Hopcroft-Ullman construction.
Note also that the Hopcroft-Ullman routine is deterministic, and consequently the construction preserves determinism.
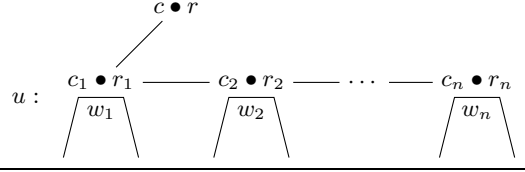
\begin{figure}
\begin{center}

\begin{tikzpicture}

\node (cr) at (5,5) {$c\bullet r$};

\foreach \i/\n in { 1/1, 3/2, 6/n}
{
\node (cr\n) at (3+\i,4) {$c_\n\bullet r_\n$};

\node (w\n) at (3+\i,3.6) {$w_\n$};

\draw (2.5+\i,3) -- (2.7+\i,3.8) -- (3.3+\i, 3.8) -- (3.5+\i,3);
}

\node (dots) at (7.5,4) {$\cdots$};
\node (u) at (3,3.8) {$u:$};
\draw (cr) -- (cr1) -- (cr2) -- (dots) -- (crn);

\end{tikzpicture}
\caption{The nested word $cc_1w_1r_1c_2w_2r_2\ldots c_nw_nr_nr$ is abstracted as a word $u$ over letters $(c_i,S_i,r_i)$ where $S_i$ is the summary of $w_i$. The position labelled by $c$ serves as initial position of the word and the corresponding state was pushed to the stack upon reading it.}\label{Figure:WordSubhedge}

\end{center}
\end{figure}
\qed
\end{proof}

\begin{theorem}\label{Thm:2DVPTrelab}
Let $A$ be a $\dtwovpt$ and $relab$ be an unambiguous letter-to-letter $\vpt$.
Then the composition $A\circ relab$ can be defined by a $\dtwovpt$.
\end{theorem}

\begin{proof}
The proof is straightforward using previous results.
First, Lemma~\ref{lemma-DecompUnambVPT} states that $relab$ can be decomposed in $T_1\circ T_2$, where $T_1$ is a deterministic $\vpt$ and $T_2$ is a co-deterministic one, and both are letter-to-letter, i.e 
$A\circ relab= A\circ T_1\circ T_2$.
Now Theorem~\ref{Thm-HUVPT} states that we can construct a $\dtwovpt$ $A'$ that realizes the composition $A\circ T_1$.
Finally, as a co-deterministic $\vpt$ can be seen as a deterministic one going right-to-left, a symmetric construction of Theorem~\ref{Thm-HUVPT} on $A'\circ T_2$ gives a $\dtwovpt$ that realizes $A\circ relab$.
\qed
\end{proof}

A look-around can be viewed as an \MSO formula with one free variable, and it is satisfied iff the formula is satisfied at this position. 
In~\cite{DBLP:conf/mfcs/MadhusudanV09}, the authors consider \MSO queries on nested words. An \MSO query is an \MSO formula with one free variable that annotates the positions of the input word that satisfies it.
They proved, using a Hopcroft-Ullman argument, that \MSO queries were also implemented by \dtwovpa.
Theorem~\ref{Thm:2DVPTrelab} proves that looks-around can be done on the fly while following the run of an other \dtwovpa. 
Since a look-around can be encoded as an unambiguous letter-to-letter
\vpt, we get the following corollary, that subsumes the result by~\cite{DBLP:conf/mfcs/MadhusudanV09}.
\begin{corollary}\label{Cor:RemovingLA}
$\dtwovpt=\dtwovptla$.
\end{corollary}

\subsection{Decision problems}

We consider the following type-checking problem: given a \vpa $A_1$ on $\Sigma$,
a finite-state automaton $A_2$ on $\Delta$, and a \dtwovpt $T$ from
$\WN{\Sigma}$ to $\Delta^*$, decide whether 
for every word $w\in \cL(A_1)$, $\inter{T}(w)$ belongs to $\cL(A_2)$.
This property is denoted by $T(A_1)\subseteq A_2$\footnote{If $A_2$ is a \vpa, the problem is known to be undecidable even for $T$ a \dvpt~\cite{DBLP:conf/icalp/RaskinS08}.}. The equivalence
problem asks whether given two \dtwovpt as input, they define the same
transduction. We prove the following result:
\begin{theorem}
\begin{enumerate}

\item The inverse image of a regular language of words by a \dtwovpt is recognizable by a \vpa.
\item The type-checking problem for \dtwovpt is \exptime-complete.
\item The equivalence problem for \dtwovpt is decidable.
\end{enumerate}
\end{theorem}

\begin{proof}
We prove the three results independently.

(1) Given a \dtwovpt $T=(A,\out)$ and an automaton on words $B$, we can define
 a \twovpa $A'$ as a product construction of $A$ and $B$ which
 simulates $B$ on the production by $\out$. States of $A'$ are simply
pairs of states of $A$ and $B$, and $A'$ recognizes  $\{w\in \WN{\Sigma} \mid \inter{T}(w)\in \cL(B)\}=\inter{T}^{-1}(\cL(B))$.
Observe that the construction is linear in the sizes of $A$ and $B$, and that as $B$ 
may be non-deterministic, $A'$ may also be non-deterministic.

(2) \exptime membership: as in the proof of the previous item, we can build
a \twovpa $A$ whose size is linear in the sizes of $T$ and $A_2$, and such that
$\cL(A)=\inter{T}^{-1}(\cL(A_2))$. Thus, $T(A_1)\subseteq A_2$
holds iff $\cL(A_1) \subseteq \cL(A)$ holds. This can be checked in \exptime
thanks to Theorem~\ref{thm:2vpa2dvpa}.

\exptime hardness: we reduce the problem of emptiness of a \dtwovpa $A$.
From $A$, we build a \dtwovpt $T=(A,\out)$ such that $\out$ maps every transition of $A$
to the empty word $\epsilon$. Then, we let $A_1$ be a \vpa  such that $\cL(A_1)=\WN{\Sigma}$
and $A_2$ such that $\cL(A_2)=\emptyset$. Then $T(A_1)\subseteq A_2$
holds iff $\cL(A)=\emptyset$.

(3) As proved in Section~\ref{Section:Expressiveness}, \dtwovpt are included in the
class of deterministic hedge-to-string transducers with look-ahead, \emph{i.e.}
deterministic top-down tree-to-string transducers with look-ahead, run on the
first-child-next-sibling encoding of the input hedge. 
The equivalence problem for these machines has recently been proven decidable 
in~\cite{SMK-FOCS15}.\qed
\end{proof}

\section{Expressiveness of Two-Way Visibly Pushdown Transducers}\label{Section:Expressiveness}

In this section, we study the expressiveness of \dtwovpt by comparing
them with Courcelle's MSO-transductions casted to nested words, the one-way model of
\cite{DBLP:conf/icalp/AlurD12}, and a top-down model for hedges, inspired by
top-down tree-to-string transducers. 

\subsection{MSO-definable Transductions}

We first define MSO for nested words and words, as done in
\cite{journals/jacm/AlurM09}, and then MSO-transductions from nested
words to words, based on Courcelle's MSO-definable graph transductions
\cite{Cour94}.

\paragraph{MSO on nested words and words} Let $\Sigma$ be a structured
alphabet. A nested word $w\in\WN{\Sigma}$ is viewed as a structure with
$\pos(w)$ as domain, over the successor predicate $S(x,y)$ interpreted as pairs $(i,i+1)$ for
$i\in\pos(w)\backslash \{|w|\}$, the label predicates $\sigma(x)$ for $\sigma\in\Sigma$,
interpreted by the positions labeled by $\sigma$, and the matching predicate
$M(x,y)$ interpreted as the set of matching pairs in $w$.

Monadic second-order logic (MSO) extends first-order logic with
quantification overs sets. First-order variables $x,y,\dots$  are
interpreted by positions of words, while second-order variables
$X,Y,\dots$ are interpreted by sets of positions.
\emph{MSO formulas for nested words over $\Sigma$} are defined by the following grammar:
{\small
$$
\varphi\ ::=\ \sigma(x)\mid x\in X\mid S(x,y)\mid M(x,y)\mid \neg
\varphi\mid \varphi\vee\varphi\mid \exists x.\varphi\mid \exists X.\varphi
$$
}
where $\sigma\in\Sigma$. 
The semantics of an MSO formula is defined in a classical way, and for
$\varphi$ an MSO formula, $w\in\WN{\Sigma}$, $\nu$ a valuation of the 
free variables of $\varphi$ into positions and sets of positions of
$w$, we write $w,\nu\models \varphi$ to mean that $w$ is a model of
$\varphi$ under the valuation $\nu$. When $\varphi$ is a sentence, we
just write $w\models \varphi$. We denote by $\MSONW[\Sigma]$ the set
of MSO formulas for nested words over $\Sigma$ (and just $\MSONW$ when
$\Sigma$ is clear from the context).
Since we are interested in transductions from nested words to words,
we also define MSO for words. Similarly as nested words, words are
seen as structures but in that case we do not have the matching pair
predicate $M(x,y)$. MSO formulas on words are defined accordingly to this smaller signature.

\begin{example}\label{ex:mso}
    We interpret $\MSONW[\Sigma]$ on nested words rather that on words
    in $\Sigma^*$. It is not a
    restriction since checking whether a given relation $M(x,y)$ is a valid matching relation is definable by an MSO formula $\phi_{wn}$.
	 This formula expresses that $M$ is a bijection
    between call and return symbols, and that it is well-nested (there is
    no crossing), as follows:
    {\small
    $$
    \begin{array}{ll}
    \neg\exists x_c,x_r,y_c,y_r. M(x_c,x_r)\wedge M(y_c,y_r)\wedge x_c\prec
    y_c\prec x_r\prec y_r \\
     \wedge\  \text{bij}(M) \wedge \forall x,y.M(x,y)\rightarrow x\prec y
    \end{array}
    $$
    }
    where $\prec$ is the transitive closure of $S$ (well-known to be
    MSO-definable) and $\text{bij}(M)$ expresses that $M$ maps
    bijectively call and return symbols (it is trivially
    MSO-definable).  
\end{example}


\paragraph{MSO transducers from nested words to words} 
MSO-transducers define (partial) functions
from nested words to word structures. The output word structure is defined by taking a
fixed number $k$ of copies of the input structure domain. Nodes of 
these copies can be filtered out by \MSONW formulas with one free first-order
variable. In particular, the nodes of the $c$-th copy are the input positions
that satisfy some given \MSONW formula $\phi_{pos}^c(x)$. The label predicates $\sigma(x)$ and the successor
predicate $S(x,y)$ of the output
structure are defined by \MSONW formulas with
respectively  one and two free first-order variables, interpreted over
the input structure. %
Formally, \emph{an MSO-transducer from nested words to words} is a
tuple $T=$
$$
(k, \phi_{dom}, (\phi_{pos}^c(x))_{1\leq c\leq k},
(\phi_{\sigma}^c(x))_{\!\!\!\!{\tiny\begin{array}{l}1{\leq} c{\leq}
                                      k\\
                                      \sigma{\in}\Sigma\end{array}}},
                              (\phi_{S}^{c,d}(x,y))_{1\leq c,d\leq k})
$$
where $k\in\mathbb{N}$ and the formulas $\phi_{dom}$,
$\phi_{pos}^c$, $\phi_{a}^c$ and
$\phi_S^{c,d}$  are
\MSONW formulas. We denote by \MSONWtoW the class of MSO-transducers
from nested words to words. 

An MSO-transducer $T$ defines a function
from nested word structures over $\Sigma$ to word structures over
$\Sigma$, denoted by
$\inter{T}$. The domain of
$\inter{T}$ consists of all nested word structures $u$ such that $u \models
\phi_{dom}$. Given a nested word structure $u\in dom(\inter{T})$, the \textit{output
  structure} $v$ such that $(u,v)\in \inter{T}$ is defined by the
 domain $D^v \subseteq \pos(u)\times \{ 1,\dots,k\}$ such that 
 $D^v = \{ (i, c)\ |\ i\in \pos(u),\ c\in\{1,\dots,k\},\
      u\models \phi_{pos}^c(i)\}$, a node $(i,c)\in D^v$ of the output
      structure is labeled $a\in\Sigma$ if  $u\models
      \phi_{a}^{c}(i)$, and a node $(j,d)\in D^v$ is the successor of
      a node $(i,c)\in D^v$ if $u\models \phi_{S}^{c,d}(i,j)$.
Note that the output structure is not necessarily a word, because for
instance, nothing guarantees that an output node is labeled by a
unique symbol, or that the successor relation forms a linear order on
the positions. However, it is not difficult to see that it is
decidable whether an \MSONWtoW transducer produces only words (see for
instance \cite{conf/icla/Filiot15}).

We say that a function $f$ from nested words to words
is MSO-definable if there exists an $T\in\MSONWtoW$ such that
$\inter{T} = f$. By definition of \MSONWtoW transducers, for any
MSO-definable function $f$ there exists $k\in \mathbb{N}$ such that
for all $u\in\dom(f)$, $|f(v)|\leq k.|u|$ (by taking $k$ as the number
of copies of the \MSONWtoW transducer defining $f$). We say in that
case that $f$ is of \emph{linear-size increase}.

\begin{example}
    This example transforms a nested word into the sequence
    of calls of maximal depth (the leaves). E.g., $c_1c_2r_2
    c_3c_4r_4r_3r_1$ is mapped to $c_2c_4$. This transformation is
    MSO-definable. The domain is defined by the formula $\phi_{wn}$
    (see Example~\ref{ex:mso}). One needs only one copy of the input
    word, whose positions are filtered out by the formula 
    $\phi_{pos}^1(x) = \exists y.M(x,y)\wedge S(x,y)$ which holds true
    iff $x$ is a call position and its successor position $y$ is its
    matching return position. The labels are
    preserved: $\phi_a^1(x) = a(x)$ for all $a\in \Sigma$. Finally,
    the successor relation is defined by 
    $\phi_S^{1,1}(x,y) = \phi_{pos}^1(x) \wedge \phi_{pos}^1(y)\wedge
    x\prec y\wedge \neg\exists z.\phi_{pos}^1(z)\wedge x\prec z
    \prec y$.

\end{example}

\subsection{Logical equivalences}

An \MSONWtoW $T$ is said to be \emph{order-preserving} if for any word $u$ of the domain of $T$,
any positions $i,j$ of $u$ and any copies $c,d$ of $T$, if $u\models \phi_S^{c,d}(i,j)$ then $i\leq j$.
This means that the output arrows can not point to the right.
It is emphasized by the next theorem, which echoes a similar result on words proved in~\cite{DBLP:conf/icalp/Bojanczyk14,conf/icla/Filiot15}.

\begin{restatable}{theorem}{ThmorderPreserv}
An order-preserving transduction is definable in \MSONWtoW 
if, and only if, it is definable by a functional\footnote{Within the
  class of \VPT, the class of
  functional \VPT is decidable in \textsf{PTime}~\cite{conf/mfcs/FiliotRRST10}} \VPT.
\end{restatable}

In the following, we show that \dtwovpt are strictly more expressive
than \MSONWtoW, and define a restriction that capture exactly \MSONWtoW.
The fact that \dtwovpt are more expressive than \MSONWtoW can
be easily shown, based on a similar result for ranked trees
established in \cite{journals/eatcs/CourcelleE12}. Since \dtwovpt can,
using their stack, express transductions of exponential-size increase,
while MSO-transductions are of linear-size increase, they are strictly
more expressive than \MSONWtoW. %

To capture exactly \MSONWtoW, one defines the \emph{single-use
  restriction} for \dtwovpt (and \dtwovptla). Intuitively, this
restriction requires that when a \dtwovpt passes twice
at the same position with the same state, 
then necessarily the transitions fired from these states produces $\epsilon$. 

\begin{definition}[Single-use restriction]
    A \dtwovpt (resp. \dtwovptla) $T=(A,\out)$ with $A =
    (Q,q_I,F,\Gamma,\delta)$ a \twovpa (resp. \twovpala) is \emph{single-use}
    with respect to a set $P\subseteq Q$ if 
    any transition $t$ from
    a state $q\not\in P$ satisfies $\out(t)=\epsilon$, and if 
    for all runs $r =
    (q_0,i_0,d_0,\sigma_0)\dots (q_\ell,i_\ell,d_\ell,\sigma_\ell)$ of $T$ on
    a word $w$
    and all states $p\in P$,
    $r$ does not visit twice the same position in state $p$, i.e. 
    if $(q_\alpha,i_\alpha) = (q_\beta,i_\beta)$ for $\alpha\neq \beta$, then $q_\alpha =
    q_\beta\not\in P$.

    A \dtwovpt (resp. \dtwovptla) is \emph{single-use}
    if it is single-use w.r.t. some set $P\subseteq Q$, and
    \emph{strongly single-use} if it is single-use w.r.t. $Q$. 
\end{definition}

We denote by $\dtwovptsu$ (resp. $\dtwovptlasu$) the class of single-use
$\dtwovpt$ (resp. $\dtwovptla$). 
By reduction to the \dtwovpa emptiness, we get:
\begin{restatable}{proposition}{PropDecidSU}
Deciding the single use property on a $\twovpt$ is \exptimecomplete.
\end{restatable}

In~\cite{journals/eatcs/CourcelleE12}, a single-use restriction was already defined for deterministic tree-walking
transducers with look-around to capture MSO-transductions from trees
to trees (and words). It requires that in any accepting run, every
node is visited at most once by a state. It is therefore more
restrictive than our single-restriction and, as a matter of fact,
corresponds to what we call the strongly single-use restriction. 
However, the following result shows that the strongly single-use restriction is not
powerful enough, in our context, to capture all MSO-definable
transductions, even with regular look-arounds. 

\begin{restatable}{lemma}{lemsula}\label{lem:sula}
    There is an MSO-definable nested word to word transduction
    $f$ which is not definable by strongly single-use \dtwovptla.
\end{restatable}

We now proceed to the first logical equivalence, between our model and
MSO-transductions, which is mainly a consequence of results from
\cite{journals/eatcs/CourcelleE12}.

\begin{restatable}{theorem}{thmMSOtDVPTLA}\label{thm:mso2dvptla}
    Let $f$ be a transduction from nested words to words. 
    Then $f$ is MSO-definable iff it is definable by a (look-around)
    \dtwovptsu, i.e., \begin{center}$\MSONWtoW = \dtwovptlasu = \dtwovptsu$.\end{center}
\end{restatable}

\begin{Sproof}
We show that both other models are equivalent to $\dtwovptlasu$.
We have
already seen that look-around can be removed from $\dtwovptla$
(Theorem~\ref{Cor:RemovingLA}), while preserving their expressive
power. Our Hopcroft-Ullman's construction can add exponentially more
visits to the same positions, but these visits are only
$\epsilon$-producing. In other words, our Hopcroft-Ullman's
construction does not preserve the strongly single-use restriction,
but it preserves the single-use restriction. As a consequence of this
observation and Corollary~\ref{Cor:RemovingLA}, we
obtain that $\dtwovptsu=\dtwovptlasu$.

To show $\MSONWtoW \subseteq \dtwovptlasu$,
we rely on the equivalence of \cite{journals/eatcs/CourcelleE12} between 
deterministic binary tree to word walking transducers with look-around
(DTWT$^{la}$) and MSO-transductions from binary trees to words
(\MSOTtoW). Informally, DTWT$^{la}$ can follow the directions of
binary trees (1st child, 2nd child and parent) and take their
transitions based on regular look-around information. Due to
determinism, they are always strongly single-use, in the sense that 
any position is not visited twice by the same state. Such a machine, running on
first-child next-sibling encoding of nested words, is easily encoded
into an equivalent \dtwovptlasu. In this encoding, a nested word over
$\Sigma$ is encoded as a binary tree over $(\Sigma_c\times
\Sigma_r)\cup \{\bot\}$, inductively defined as
$\fcns(cw_1rw_2) = (c,r)(\fcns(w_1),\fcns(w_2))$ and $\fcns(\epsilon)
= \bot$. In this encoding, moving to a 1st child corresponds to
moving from $c$ to $w_1$, which can be done by a $\dtwovptlasu$, and moving to a
2nd child corresponds to moving from $c$ to $w_2$. This can be done
also by a $\dtwovptlasu$, but it needs to traverse all the word
$cw_1r$, while producing $\epsilon$ only. Similarly, one can encode
moves to parent nodes. The two latter moves implies that the
$\dtwovptlasu$ is not strongly single-use anymore, but it remains
single-use: the extra moves are all $\epsilon$-producing. The result
follows as $\MSONWtoW = \MSOTtoW\circ \fcns$.

To show $\dtwovptlasu \subseteq \MSONWtoW$, we rely on another
correspondence shown in  \cite{journals/eatcs/CourcelleE12}, between 
\MSOTtoW and deterministic (visibly) pushdown binary tree to word
walking transducers with look-around of linear-size increase
(DPTWT$_{lsi}^{la}$). These transducers extend DTWT$^{la}$ with a
pushdown store with a visibly condition: when moving to a child, they
push one symbol, and moving up, they pop one symbol. The lsi
restriction is semantical: they restrict the class to transducers that
define lsi transductions. Any $\dtwovptlasu$ defines an lsi
transduction, and can be easily encoded into a DPTWT$_{lsi}^{la}$
running on \fcns encodings, which mimics the moves of the
$\dtwovptlasu$. Again, the result follows by the equality $\MSONWtoW =
\MSOTtoW \circ \fcns$.\qed
\end{Sproof}

\subsection{Comparison with other transducer models}

In this section, we relate \dtwovpt to two other transducer models, namely
streaming tree-to-string transducers and deterministic hedge-to-string transducers
with look-ahead. Streaming tree-to-string transducers with a simple
copyless restriction of updates will serve as the third edge of our
trinity. Deterministic hedge-to-string transducers with look-ahead is
a natural model for which equivalence is known to be decidable.

Streaming tree-to-string transducers are deterministic one-way
machines \cite{DBLP:conf/icalp/AlurD12} equipped with registers
storing words. 
We fix a finite alphabet $\Delta$ and, given two finite sets $\cal X$ and $\cal Y$, denote by 
$\upd(\mathcal{X},\mathcal{Y})$ the set of mappings from $\mathcal{X}$ to  $(\Delta \cup \mathcal{Y})^*$.


\begin{definition}
  A \emph{streaming tree-to-string transducer} $S$ (\stst for short) is a
  deterministic machine defined over a structured alphabet $\Sigma$
  and given by the tuple $(Q,q_I,\Gamma,\mathcal{X},\delta,\mu_F)$ where
  $Q$ is a finite set of states, $q_I \in Q$ is the initial state,
  $\Gamma$ is a finite set of stack symbols and $\mathcal{X}$ is a
  finite set of registers. Finally, $\mu_F$ is a partial mapping from
  $Q$ to $(\Delta \cup \mathcal{X})^*$ and $\delta=\delta^{push} \uplus
  \delta^{pop}$  where
  $\delta^{push} : Q \times \Sigma_c \to Q \times \Gamma \times \upd(\mathcal{X},\mathcal{X})$
and $\delta^{pop} : Q \times \Sigma_r \times \Gamma \to Q \times
\upd(\mathcal {X},\mathcal{X}\cup \mathcal{X}')$, $\mathcal{X}'$ being a
disjoint copy of $\mathcal{X}$.   
\end{definition}

Let $\mathcal{V}^\Delta_\mathcal{X}$ be the set of mappings from from $\mathcal{X}$ to
  $\Delta^*$. These mappings are extended to $(\mathcal{X}\cup\Delta)^*$ by
considering them as identity over $\Delta$.
An accepting run of a \stst $S$ on a nested word $w$ is a
(non-empty) sequence $(q_0,\theta_0,\sigma_0, w_0) \ldots
(q_\ell,\theta_\ell,\sigma_\ell, w_\ell)$ of
quadruples from $Q \times \mathcal{V}^\Delta_\mathcal{X} \times 
(\Gamma \times \mathcal{V}^\Delta_\mathcal{X})^*  \times \Sigma^*$ such that 
$q_0=q_I$, $w_0=w$, $w_\ell=\epsilon$, $\theta_0$ is the
mapping $\theta_\epsilon$ which associates $\epsilon$ to any $X$ in
$\mathcal{X}$, $\sigma_0$, $\sigma_\ell$ are equal to $\bot$ the empty
stack and for all $0 \le i < \ell$, one has either 
\begin{itemize}
\item $w_i = c w_{i+1}$ and  there exists $(q_i,c,q_{i+1}, \gamma,
  \nu) \in \delta^{push}$, $\theta_{i+1}=\theta_\epsilon$ and
  $\sigma_{i+1}=\sigma_i(\gamma,\theta_i \circ \nu)$, 
\item $w_i = r w_{i+1}$ and there exists $(q_i,r,\gamma,q_{i+1}, 
  \nu) \in \delta^{pop}$, $\sigma_i=\sigma_{i+1}(\gamma,\theta)$
and $\theta_{i+1}=\theta' \circ \theta_i \circ \nu$, where $\theta'\in \mathcal{V}^\Delta_\mathcal{X'}$
is defined by $\theta'(X')=\theta(X)$ for all $X\in \mathcal{X}$.
\end{itemize} 

The semantics $\sem{S}$ of the \stst $S$ is a partial mapping from $\WN{\Sigma}$
  to $\Delta^*$ such that $\sem{S}(w)=v$ if there exists an accepting
  run on $w$ in $S$ ending in some configuration $(q_\ell,\theta_\ell,\bot, \epsilon)$
 and $v = \theta_{\ell}(\mu_F(q_{\ell}))$. 

Using a restriction on the updates $\upd$ used in \stst (so-called 
copyless updates), \cite{DBLP:conf/icalp/AlurD12} proved
that copyless \stst and  \MSONWtoW are expressively equivalent.
As a consequence, we obtain the logic/two-way/one-way trinity
announced in the introduction:
\begin{theorem}
$
\MSONWtoW = \text{\biDVPTSU} = \textup{copyless }\stst
$
\end{theorem}

A well-known class of transducers running on ranked trees is the class
of deterministic top-down tree transducers with look-ahead. This class can be
defined to output strings. We consider now the extension of this class
to unranked trees, or more precisely sequences of unranked trees, that
is, hedges.

\begin{definition}
  An \emph{hedge automaton} (\Ha for short) over the structured
  alphabet $\Sigma$~\footnote{Usually, such automata are given over a
    classical unstructured but unary alphabet. However, for having a uniform
    presentation, we choose wlog this definition which corresponds
    somehow to consider a pair from $\Sigma_c \times \Sigma_r$ as
    single symbol.} is a tuple $(Q,F,\delta)$ where $Q$ is a finite
  set of states, $F \subseteq Q$ is a set of final states and $\delta$
  is a transition relation such that $\delta \subseteq Q \times \Sigma_c \times
  \Sigma_r \times Q \times Q$.
\end{definition}

An hedge automaton is said to be bottom-up deterministic if whenever
$(q,c,r,q_1,q_2)$ and $(q',c,r,q_1,q_2)$ belongs to $\delta$, it holds
that $q=q'$. 
The semantics of an \Ha $B$ is given by means of sets $\cL^B_q \subseteq
\WN{\Sigma}$ defined for each $q \in Q$ inductively as follows: $(i)$
$\epsilon \in \cL^B_q$ for all $q$ and $(ii)$ $cwrw' \in \cL^B_q$ if
$(q,c,r,q_1,q_2) \in \delta$ and $w \in \cL^B_{q_1}$, $w' \in
\cL^B_{q_2}$. The language defined by an \Ha $B$ is then $\bigcup_{q \in
  F} L^B_q$. Note that when $B$ is bottom-up deterministic whenever
$q_1 \neq q_2$, it holds that $\cL^B_{q_1} \cap \cL^B_{q_2} =
\varnothing$.

\begin{definition}
  A deterministic \emph{hedge-to-string transducer} with look-ahead
  (\dHtoSla) $H$ over the structured alphabet $\Sigma$ and the output
  alphabet $\Delta$ is given by a tuple $(Q,I,F,\delta,B)$ where $Q$
  is a finite set of states, $q_I \in Q$ is an initial state, $F
  \subseteq Q$ is a set of final states, $B$ is a deterministic
  bottom-up hedge automaton with states $Q'$, and $\delta$ is a
  transition relation given by a partial mapping
$$\delta : Q \times \Sigma_c \times \Sigma_r \times Q' \times Q' \to
\Delta_Q^*$$
$\Delta_Q$ is the finite set of symbols $(\Delta
\cup \{q(x_i) \mid 1 \le i \le 2, q \in Q\})^*$. 
\end{definition}

The semantics of a \dHtoSla is first given by a partial mapping
$\sem{H}$ from $\WN{\Sigma} \times  Q$ onto $\Delta^*$ defined inductively as:
$(i)$ $\sem{H}(\epsilon,q)=\epsilon$ if $q \in F$, and $(ii)$ for $w=c w_1 r w_2$
with $w_1,w_2 \in \WN{\Sigma}$, $\sem{H}(w,q) = \omega\lbrack q_i(x_{i_j}) \leftarrow
  \sem{H}(w_{i_j},q_i) \rbrack$ where $\omega\lbrack q_i(x_{i_j}) \leftarrow
  \sem{H}(w_{i_j},q_i) \rbrack$ denotes the word $\omega$ in which each
  occurrence of $q_i(x_{i_j})$ has been replaced by
  $\sem{H}(w_{i_j},q_i)$ if  $\delta(q,c,r,q',q'')=\omega$, $w_1 \in
  \cL^{q'}_B$ and $w_2 \in \cL^{q''}_B$ and undefined otherwise. 

Then, the transduction $\sem{H}$ defined by $H$ is given by $\{(w,s) \mid w \in \WN{\Sigma}, \ s=\sem{H}(w,q_I)\}$.



 \begin{restatable}{theorem}{EquivModels}
 $\dtwovpt \subsetneq \stst$ and $\dtwovpt \subsetneq \dHtoSla$
 \end{restatable}

\begin{Sproof} The two results rely on a same intermediate
model that extends the transition algebra described in Section~\ref{Section:Twovpa}.
This algebra allows to describe the possible traversals of a \dtwovpa.
One can extend it to \dtwovpt by storing in matrices the words
produced by traversals. This yields an infinite algebra, realized
by a finite set of operations. We use this to describe
effective translations into \stst and \dHtoSla.

As an illustration, in order to build of an equivalent
\stst, the set of variables considered is the set 
$\Xi=\{
  x^{(p,d),(p',d')} \mid (p,d),(p',d') \in Q \times \Moves\}$, \emph{i.e.}
one variable for each traversal. This generalizes the construction
described in~\cite{alur_et_al:LIPIcs:2010:2853,DBLP:conf/lics/AlurFT12} 
in order to translate a deterministic two-way transducer
(on words) into a streaming string transducer.

The fact that the inclusions are strict relies on a simple argument
based on size increase: on nested words of bounded depth, 
\dtwovpt are linear-size increase, while \stst and
\dHtoSla are not.
\qed
\end{Sproof}





\section{Discussion}

\paragraph{Unranked tree to word transductions} Since unranked trees $t$
can be linearised into nested words $lin(t)$, our result also gives a model for
unranked tree to word transductions. If one denotes by $\MSOUtoW$ the
transductions from unranked trees to words definable by an MSO
transducer (over the signature of unranked trees that has the child
and next-sibling predicates), it is easy to show that $\MSOUtoW =
\dtwovptsu \circ lin.$

One could argue that \dtwovpt for realising transductions of
unranked trees is not an adequate
model, because it performs unnecessary
$\epsilon$-producing moves to navigate, for instance, from a node $n$
to its next-sibling. Indeed, the \dtwovpt needs to walk through the whole subtree
rooted at $n$.

First, while it is true from an operational point of view, we
think that the simplicity of \dtwovpt makes them a good candidate as a
specification model of unranked tree transductions, and to this aim,
it is easy to define, as we did for next-sibling moves (rules 
$q\xrightarrow{(c,r)} p$), macros that
realise moves given by the predicates of unranked trees (and their
inverse).  
Second, for instance in the context of stream processing of XML documents, it
cannot be always assumed that the input document is given by its DOM
(with the unranked tree predicates) as sometimes, it is just stored as
plain text, i.e. as its linearisation.

Finally and most importantly,   our result allows one to get an extension of a
known model of ranked tree to word transductions, to unranked tree to
word transductions, namely, \emph{deterministic pushdown unranked tree to word
walking transducers} (DPUWT). To avoid technical details, we define formally
this model only in Appendix, and rather give intuitions here. DPUWT
can walk through the unranked tree following the next-sibling and
first-child predicates (and their inverse), while producing words on the output. 
They are also equipped with a pushdown store with
a visibly condition: whenever they go down the tree by one level, they
have to push one symbol onto the stack, and going up, they pop one
symbol. They let the stack unchanged when moving horizontally between
siblings. With the single-use restriction, defined similarly as for
\dtwovpt, we get that $\MSOUtoW = \text{DPUWT}_{su}$. Therefore,
if the input is given by an unranked tree, one can rather use a DPUWT
or a \dtwovpt on the linearisation.


\paragraph{Nested word to nested word transductions} As we claimed
earlier, \biDVPTSU can be used to define
unranked tree transformations represented as nested word to nested word
transducers, that is, as nested word to word tranduscers with a
structured output alphabet. 

On the logical side, \MSONWtoW transductions
can be extended with binary formulas $\varphi_M^{c,d}(x,y)$ aiming at representing
the matching relation existing on output nested words. As
checking whether a relation denotes a matching relation is \MSO definable 
(see Example \ref{ex:mso}), one
can decide whether any input nested word is indeed transformed by the
\MSONWtoW transducer into a nested word by testing the validity of the
sentence obtained from the logical definition of the matching $M$
(Example~\ref{ex:mso}) by
replacing the predicate $M$ with $\bigvee_{c,d}\varphi^{c,d}_M$. 
So, starting from an \MSONWtoW transducer with a matching relation defined
on its output, one may forget this matching and view this transducer
as an ordinary \MSONWtoW transducer; this machine turns out to be equivalent in
the sense that remaining call and returns symbols induce uniquely the erased
matching.  Finally, by the results presented in this paper, one can
from this \MSONWtoW transducer build an equivalent
\biDVPTSU whose range will indeed contain only nested words and thus,
defines an unranked tree transformation. 

Let us point out that our results do not entail the trinity for
tree-to-tree transformations: the class of D2VPT which produce
only nested words/trees as output may be a good candidate to complete
the missing part (the equivalence between MSO transformations and
streaming tree transducers has already been established in
\cite{DBLP:conf/icalp/AlurD12}). Nonetheless, deciding this class
seems to be challenging and moreover, there is actually no guarantee
that it corresponds to the other two cited members of this trinity.



\paragraph{Input streaming} In an input streaming scenario, one
assumes that the input nested word is given as a stream of call and
return symbols. In such a scenario, one wants to transform the input
stream as soon as possible, on-the-fly, and it is not reasonable to
load the whole stream in memory. An
interesting question is whether a given \biDVPT really needs its
two-way ability ? In other words, can we decide whether a given
\biDVPT is equivalent to a (one-way) \VPT ? For words and two-way
finite transducers, this question has been shown to be decidable in
\cite{conf/lics/FiliotGRS13}. As future work, we want to extend this
result to \biDVPT.






\bibliographystyle{plainnat}
\bibliography{papers}

\begin{thebibliography}{10}
\providecommand{\natexlab}[1]{#1}
\providecommand{\url}[1]{\texttt{#1}}
\expandafter\ifx\csname urlstyle\endcsname\relax
  \providecommand{\doi}[1]{doi: #1}\else
  \providecommand{\doi}{doi: \begingroup \urlstyle{rm}\Url}\fi

\bibitem[Alur and Madhusudan(2004)]{AM04}
Rajeev Alur and P.~Madhusudan.
\newblock Visibly pushdown languages.
\newblock In \emph{Proceedings of the 36th {A}nnual {ACM} {S}ymposium on
  {T}heory of {C}omputing}, pages 202--211 (electronic). ACM, New York, 2004.

\bibitem[Alur and Madhusudan(2009)]{journals/jacm/AlurM09}
Rajeev Alur and P.~Madhusudan.
\newblock Adding nesting structure to words.
\newblock \emph{J. ACM}, 56\penalty0 (3), 2009.

\bibitem[Comon{-}Lundh et~al.(2007)Comon{-}Lundh, Dauchet, Gilleron,
  L{\"o}ding, Jacquemard, Lugiez, Tison, and Tommasi]{TATA07}
Hubert Comon{-}Lundh, Max Dauchet, R{\'e}mi Gilleron, Cristof L{\"o}ding,
  Florent Jacquemard, Denis Lugiez, Sophie Tison, and Marc Tommasi.
\newblock \emph{Tree Automata Techniques and Applications}.
\newblock online, November 2007.
\newblock URL \url{http://www.grappa.univ-lille3.fr/tata/}.

\bibitem[Courcelle and Engelfriet(2012)]{journals/eatcs/CourcelleE12}
Bruno Courcelle and Joost Engelfriet.
\newblock Book: Graph structure and monadic second-order logic. {A}
  language-theoretic approach.
\newblock \emph{Bulletin of the EATCS}, 108:\penalty0 179, 2012.

\bibitem[Filiot(2015)]{conf/icla/Filiot15}
Emmanuel Filiot.
\newblock Logic-automata connections for transformations.
\newblock In \emph{ICLA}, volume 8923 of \emph{LNCS}, pages 30--57. Springer,
  2015.

\bibitem[Filiot and Servais(2012)]{FS12}
Emmanuel Filiot and Fr{\'{e}}d{\'{e}}ric Servais.
\newblock Visibly pushdown transducers with look-ahead.
\newblock In \emph{{SOFSEM} 2012}, pages 251--263, 2012.

\bibitem[Hopcroft and Ullman(1967)]{HopUll67}
J.~E. Hopcroft and J.~D. Ullman.
\newblock An approach to a unified theory of automata.
\newblock \emph{BELLTJ: The Bell System Technical Journal}, 46:\penalty0
  1793--1829, 1967.

\bibitem[Neven and Schwentick(2002)]{NevSch02}
Franck Neven and Thomas Schwentick.
\newblock Query automata over finite trees.
\newblock \emph{Theoretical Computer Science}, 275, 2002.

\bibitem[Niehren et~al.(2005)Niehren, Planque, Talbot, and
  Tison]{conf/dbpl/NiehrenPTT05}
Joachim Niehren, Laurent Planque, Jean-Marc Talbot, and Sophie Tison.
\newblock {N}-ary queries by tree automata.
\newblock In \emph{DBLP}, volume 3774 of \emph{LNCS}, pages 217--231. Springer,
  2005.

\bibitem[Shepherdson(1959)]{Shepherdson59}
J.~C. Shepherdson.
\newblock The reduction of two-way automata to one-way automata.
\newblock \emph{IBM Journal of Research and Development}, 3\penalty0
  (2):\penalty0 198--200, 1959.

\end{thebibliography}


\begin{thebibliography}{34}
\providecommand{\natexlab}[1]{#1}
\providecommand{\url}[1]{\texttt{#1}}
\expandafter\ifx\csname urlstyle\endcsname\relax
  \providecommand{\doi}[1]{doi: #1}\else
  \providecommand{\doi}{doi: \begingroup \urlstyle{rm}\Url}\fi

\bibitem[Alur and {\v C}ern{\'y}(2010)]{alur_et_al:LIPIcs:2010:2853}
R.~Alur and P.~{\v C}ern{\'y}.
\newblock Expressiveness of streaming string transducers.
\newblock In \emph{FSTTCS}, volume~8, pages 1--12, 2010.

\bibitem[Alur and {\v C}ern{\'y}(2011)]{conf/popl/AlurC11}
R.~Alur and P.~{\v C}ern{\'y}.
\newblock Streaming transducers for algorithmic verification of single-pass
  list-processing programs.
\newblock In \emph{POPL}, pages 599--610, 2011.

\bibitem[Alur(2016)]{webnested}
Rajeev Alur.
\newblock Nested words, 2016.
\newblock URL \url{https://www.cis.upenn.edu/~alur/nw.html}.

\bibitem[Alur and D'Antoni(2012)]{DBLP:conf/icalp/AlurD12}
Rajeev Alur and Loris D'Antoni.
\newblock Streaming tree transducers.
\newblock In \emph{ICALP (2)}, volume 7392 of \emph{LNCS}, pages 42--53.
  Springer, 2012.

\bibitem[Alur and Madhusudan(2009)]{journals/jacm/AlurM09}
Rajeev Alur and P.~Madhusudan.
\newblock Adding nesting structure to words.
\newblock \emph{J. ACM}, 56\penalty0 (3), 2009.

\bibitem[Alur et~al.(2012)Alur, Filiot, and Trivedi]{DBLP:conf/lics/AlurFT12}
Rajeev Alur, Emmanuel Filiot, and Ashutosh Trivedi.
\newblock Regular transformations of infinite strings.
\newblock In \emph{LICS}, pages 65--74, 2012.

\bibitem[Bloem and Engelfriet(2000)]{DBLP:journals/jcss/BloemE00}
Roderick Bloem and Joost Engelfriet.
\newblock A comparison of tree transductions defined by monadic second order
  logic and by attribute grammars.
\newblock \emph{J. Comput. Syst. Sci.}, 61\penalty0 (1):\penalty0 1--50, 2000.

\bibitem[Bojanczyk(2014)]{DBLP:conf/icalp/Bojanczyk14}
Mikolaj Bojanczyk.
\newblock Transducers with origin information.
\newblock In \emph{ICALP}, volume 8573 of \emph{LNCS}, pages 26--37. Springer,
  2014.

\bibitem[Chytil and J{\'a}kl(1977)]{ICALP::ChytilJ1977}
Michal Chytil and Vojtech J{\'a}kl.
\newblock Serial composition of 2-way finite-state transducers and simple
  programs on strings.
\newblock In \emph{ICALP}, volume~52 of \emph{LNCS}, pages 135--147. Springer,
  1977.

\bibitem[Comon{-}Lundh et~al.(2007)Comon{-}Lundh, Dauchet, Gilleron,
  L{\"o}ding, Jacquemard, Lugiez, Tison, and Tommasi]{TATA07}
Hubert Comon{-}Lundh, Max Dauchet, R{\'e}mi Gilleron, Cristof L{\"o}ding,
  Florent Jacquemard, Denis Lugiez, Sophie Tison, and Marc Tommasi.
\newblock \emph{Tree Automata Techniques and Applications}.
\newblock online, November 2007.
\newblock URL \url{http://www.grappa.univ-lille3.fr/tata/}.

\bibitem[Courcelle(1994)]{Cour94}
B.~Courcelle.
\newblock Monadic second-order definable graph transductions: a survey.
\newblock \emph{Theoretical Computer Science}, 126(1):\penalty0 53--75, 1994.

\bibitem[Courcelle and Engelfriet(2012)]{journals/eatcs/CourcelleE12}
Bruno Courcelle and Joost Engelfriet.
\newblock Book: Graph structure and monadic second-order logic. {A}
  language-theoretic approach.
\newblock \emph{Bulletin of the EATCS}, 108:\penalty0 179, 2012.

\bibitem[Culik and Karhumaki(1987)]{CulKar87}
K.~Culik and J.~Karhumaki.
\newblock The equivalence problem for single-valued two-way transducers (on
  {NPDT0L} languages) is decidable.
\newblock \emph{SIAM J. on Computing}, 16\penalty0 (2):\penalty0 221--230,
  1987.

\bibitem[Engelfriet and Hoogeboom(2001)]{EngHoo01}
Joost Engelfriet and Hendrik~Jan Hoogeboom.
\newblock {MSO} definable string transductions and two-way finite-state
  transducers.
\newblock \emph{ACM Transactions on Computational Logic}, 2\penalty0
  (2):\penalty0 216--254, 2001.

\bibitem[Engelfriet and Maneth(1999)]{MTT}
Joost Engelfriet and Sebastian Maneth.
\newblock Macro tree transducers, attribute grammars, and mso definable tree
  translations.
\newblock \emph{Information and Computation}, 154\penalty0 (1):\penalty0
  34--91, 1999.

\bibitem[Engelfriet and Maneth(2003)]{DBLP:journals/siamcomp/EngelfrietM03}
Joost Engelfriet and Sebastian Maneth.
\newblock Macro tree translations of linear size increase are {MSO} definable.
\newblock \emph{{SIAM} J. of Computing}, 32\penalty0 (4):\penalty0 950--1006,
  2003.

\bibitem[Filiot et~al.(2011)Filiot, Gauwin, Reynier, and
  Servais]{conf/fsttcs/FiliotGRS11}
E.~Filiot, O.~Gauwin, P.-A. Reynier, and F.~Servais.
\newblock Streamability of nested word transductions.
\newblock In \emph{FSTTCS}, volume~13 of \emph{LIPIcs}, pages 312--324, 2011.

\bibitem[Filiot(2015)]{conf/icla/Filiot15}
Emmanuel Filiot.
\newblock Logic-automata connections for transformations.
\newblock In \emph{ICLA}, volume 8923 of \emph{LNCS}, pages 30--57. Springer,
  2015.

\bibitem[Filiot et~al.(2010)Filiot, Raskin, Reynier, Servais, and
  Talbot]{conf/mfcs/FiliotRRST10}
Emmanuel Filiot, Jean-Fran{\c c}ois Raskin, Pierre-Alain Reynier,
  Fr{\'e}d{\'e}ric Servais, and Jean-Marc Talbot.
\newblock Properties of visibly pushdown transducers.
\newblock In \emph{MFCS}, volume 6281 of \emph{LNCS}, pages 355--367. Springer,
  2010.

\bibitem[Filiot et~al.(2013)Filiot, Gauwin, Reynier, and
  Servais]{conf/lics/FiliotGRS13}
Emmanuel Filiot, Olivier Gauwin, Pierre-Alain Reynier, and Fr{\'e}d{\'e}ric
  Servais.
\newblock From two-way to one-way finite state transducers.
\newblock In \emph{LICS}, pages 468--477. IEEE, 2013.

\bibitem[Gauwin et~al.(2011)Gauwin, Niehren, and
  Tison]{journals/iandc/GauwinNT11}
O.~Gauwin, J.~Niehren, and S.~Tison.
\newblock Queries on {XML} streams with bounded delay and concurrency.
\newblock \emph{Information and Computation}, 209\penalty0 (3):\penalty0
  409--442, 2011.

\bibitem[Gurari(1982)]{Gurari82}
Eitan Gurari.
\newblock The equivalence problem for deterministic two-way sequential
  transducers is decidable.
\newblock \emph{SIAM J. on Computing}, 11, 1982.

\bibitem[Hopcroft and Ullman(1967)]{HopUll67}
J.~E. Hopcroft and J.~D. Ullman.
\newblock An approach to a unified theory of automata.
\newblock \emph{BELLTJ: The Bell System Technical Journal}, 46:\penalty0
  1793--1829, 1967.

\bibitem[Kumar et~al.(2007)Kumar, Madhusudan, and
  Viswanathan]{conf/www/KumarMV07}
Viraj Kumar, P.~Madhusudan, and Mahesh Viswanathan.
\newblock Visibly pushdown automata for streaming {XML}.
\newblock In \emph{WWW}, pages 1053--1062. ACM, 2007.

\bibitem[Madhusudan and Viswanathan(2009)]{DBLP:conf/mfcs/MadhusudanV09}
P.~Madhusudan and Mahesh Viswanathan.
\newblock Query automata for nested words.
\newblock In \emph{MFCS}, volume 5734 of \emph{LNCS}, pages 561--573. Springer,
  2009.

\bibitem[Neven and Schwentick(2002)]{NevSch02}
Franck Neven and Thomas Schwentick.
\newblock Query automata over finite trees.
\newblock \emph{Theoretical Computer Science}, 275, 2002.

\bibitem[Niehren et~al.(2005)Niehren, Planque, Talbot, and
  Tison]{conf/dbpl/NiehrenPTT05}
Joachim Niehren, Laurent Planque, Jean-Marc Talbot, and Sophie Tison.
\newblock {N}-ary queries by tree automata.
\newblock In \emph{DBLP}, volume 3774 of \emph{LNCS}, pages 217--231. Springer,
  2005.

\bibitem[P{\'{e}}cuchet(1985)]{Pecuchet85}
Jean{-}Pierre P{\'{e}}cuchet.
\newblock Automates boustrophedon, semi-groupe de {B}irget et monoide inversif
  libre.
\newblock \emph{{RAIRO - ITA}}, 19\penalty0 (1):\penalty0 71--100, 1985.

\bibitem[Picalausa et~al.(2011)Picalausa, Servais, and
  Zim{\'a}nyi]{conf/sac/PicalausaSZ11}
Fran{\c c}ois Picalausa, Fr{\'e}d{\'e}ric Servais, and Esteban Zim{\'a}nyi.
\newblock {XE}volve: an {XML} schema evolution framework.
\newblock In \emph{SAC}, pages 1645--1650. ACM, 2011.

\bibitem[Raskin and Servais(2008)]{DBLP:conf/icalp/RaskinS08}
Jean{-}Fran{\c{c}}ois Raskin and Fr{\'{e}}d{\'{e}}ric Servais.
\newblock Visibly pushdown transducers.
\newblock In \emph{ICALP}, volume 5126 of \emph{LNCS}, pages 386--397.
  Springer, 2008.
\newblock \doi{10.1007/978-3-540-70583-3_32}.
\newblock URL \url{http://dx.doi.org/10.1007/978-3-540-70583-3_32}.

\bibitem[Segoufin and Sirangelo(2007)]{conf/icdt/SegoufinS07}
L.~Segoufin and C.~Sirangelo.
\newblock Constant-memory validation of streaming {XML} documents against
  {DTD}s.
\newblock In \emph{ICDT}, volume 4353 of \emph{LNCS}, pages 299--313. Springer,
  2007.

\bibitem[Seidl et~al.(2015)Seidl, Maneth, and Kemper]{SMK-FOCS15}
Helmut Seidl, Sebastian Maneth, and Gregor Kemper.
\newblock Equivalence of deterministic top-down tree-to-string transducers is
  decidable.
\newblock In \emph{FOCS}, pages 943--962. {IEEE}, 2015.

\bibitem[Shepherdson(1959)]{Shepherdson59}
J.~C. Shepherdson.
\newblock The reduction of two-way automata to one-way automata.
\newblock \emph{IBM Journal of Research and Development}, 3\penalty0
  (2):\penalty0 198--200, 1959.

\bibitem[Thomas(1997)]{Tho97handbook}
W.~Thomas.
\newblock Languages, automata and logic.
\newblock In A.~Salomaa and G.~Rozenberg, editors, \emph{Handbook of Formal
  Languages}, volume 3, Beyond Words. Springer, Berlin, 1997.

\end{thebibliography}

\newpage

\appendix
\section{Appendix}

\subsection{Two-way visibly pushdown automata}

\propSimCongruence*

\begin{proof}
We consider $R$ the set of binary relations over $Q \times \Moves$.
Obviously, $R$ is finite. As traversals are subsets of $R$, $\sim$ is
of finite index.
Let us now prove that $\sim$ is a congruence relation for the binary
operation $.$ and the unary ones, $f_{c,r}$ (for $c \in \Sigma_c, r
\in \Sigma_r$).

From $\sim$, we define four equivalence relations $\sim_{\ll}, \sim_{\lr},
\sim_{\rl}, \sim_{\rr}$ on $Q \times Q$ such that  for $(\alpha,\beta \in
\{\mathsf{l},\mathsf{r}\} )$, we have $u \sim_{\alpha\beta} v$ if
\begin{align*}
&\lbrack u\rbrack_{\sim} \cap (Q
\times  \{bdir(\alpha)\}
\times Q \times \{edir(\beta)\})   = \\
&\lbrack v\rbrack_{\sim} \cap (Q
\times  \{bdir(\alpha)\}\times Q \times \{edir(\beta)\})  
\end{align*} 
where $bdir(\mathsf{l}) = edir(\mathsf{r})=\rmove$ and $bdir(\mathsf{r}) = edir(\mathsf{l})=\lmove$.

Intuitively, $(p,q)$ belongs to $\lbrack{w}\rbrack_{\sim_{\ll}}$
(respectively to $\lbrack{w}\rbrack_{\sim_{\lr}}$) if there exists a
run of $A$ on $w$ starting reading $w$ from the \emph{left} side,
\textsl{ie}, with direction $\rmove$ in state $p$ and leaves the word
on the \emph{left}, \textsl{ie}, with direction $\lmove$ (resp. on the
\emph{right}, \textsl{ie}, with direction $\rmove$) in state $q$.

The relation $\sim$ is uniquely determined by the four relations
$\sim_{\ll}, \sim_{\lr}, \sim_{\rl}, \sim_{\rr}$ and in particular, $\sim$
is a congruence iff all the $\sim_{\ll}, \sim_{\lr}, \sim_{\rl},
\sim_{\rr}$ are congruences.

Let us first notice that for $\epsilon$, one has
$\eqclass{\epsilon}{\sim_{\ll}}=\eqclass{\epsilon}{\sim_{\rr}}=\varnothing$
whereas $\eqclass{\epsilon}{\sim_{\lr}}$,
$\eqclass{\epsilon}{\sim_{\rl}}$ are the identity relation.

Let us consider $u,u',v,v'$ in $\WN{\Sigma}$ and assume that 
$u \sim u'$ (and thus, $ u \sim_{\ll} u'$, $ u \sim_{\lr} u'$, $u
\sim_{\rl} u'$, $u \sim_{\rr} u'$) and $v \sim v'$. We consider $u.v$
and $u.'v'$ and prove that $u.v \sim u.'v'$.  

From the definition of runs and traversals, one has 


$$
\begin{array}{l}
\eqclass{u.v}{\sim_{\ll}} = \eqclass{u}{\sim_{\ll}}  \cup
\eqclass{u}{\sim_{\lr}} \circ (\eqclass{v}{\sim_{\ll}} \circ
\eqclass{u}{\sim_{\rr}})^* \circ 
\eqclass{v}{\sim_{\ll}}  \circ  \eqclass{u}{\sim_{\rl}} \\ 
\eqclass{u.v}{\sim_{\lr}} = \eqclass{u}{\sim_{\lr}}  \circ ( 
\eqclass{v}{\sim_{\ll}} \circ
\eqclass{u}{\sim_{\rr}})^* \circ 
\eqclass{v}{\sim_{\lr}} \\ 
\eqclass{u.v}{\sim_{\rl}}= \eqclass{v}{\sim_{\rl}} \circ (
\eqclass{u}{\sim_{\rr}}  \circ \eqclass{v}{\sim_{\ll}})^* \circ 
\eqclass{u}{\sim_{\rl}} \\ 
\eqclass{u.v}{\sim_{\rr}} = \eqclass{v}{\sim_{\rr}} \cup 
\eqclass{v}{\sim_{\rl}}
\circ ( \eqclass{u}{\sim_{\rr}} \circ
\eqclass{v}{\sim_{\ll}})^* \circ 
\eqclass{u}{\sim_{\rr}} \circ  \eqclass{v}{\sim_{\lr}}
\end{array}
$$

Hence, $\eqclass{u.v}{\sim_{\alpha\beta}} =
\eqclass{u'.v'}{\sim_{\alpha\beta}}$ for all $\alpha,\beta \in \{\mathsf{l},\mathsf{r}\}$
and so, $u.v \sim u'.v'$.
Let us point out that these definition are similar to those defined
for words in the case of two-way finite state automata \citeappendix{Shepherdson59} and
that
$\eqclass{(u.v).w}{\sim_{\alpha\beta}}=\eqclass{u.(v.w)}{\sim_{\alpha\beta}}$
and $\eqclass{u.\epsilon}{\sim_{\alpha\beta}} =
\eqclass{\epsilon.u}{\sim_{\alpha\beta}}$ for all $\alpha,\beta \in
\{\mathsf{l},\mathsf{r}\}$.

Now, let us consider $u,u'$ in $\WN{\Sigma}$ and assume that $u \sim
u'$. We consider $cur=f_{c,r}(u)$ and $cu'r = f_{c,r}(u')$ and show
that $cur \sim cu'r$.  Expressing traversals on $cur$ is much more
intricate. To ensure that traversals abstract properly runs, we need
to forget about stack contents and thus, reason again only on
nested words when composing sub-runs of $cwr$. Hence, new
notations are needed: we let for $d$ in $\{\lmove,\rmove\}$ and $w \in
\WN{\Sigma}$
{\small
$$ 
Z^{c}_{\ll,d} = 
\left \{  (p,q) \mid 
(p,\rmove,c,p',\lmove,\gamma) \in \delta^{\text{push}} \ , \ 
(p',\lmove,\gamma,c,q,d) \in \delta^{\text{pop}}
\right \}
$$
%
%
$$Z^{cw}_{\ll,d} = \bigcup_{\gamma \in \Gamma}  
\left ( \bigcup_{(\overrightarrow{p}\xrightarrow{c,+\gamma}\overrightarrow{p\prime})}  \{(p,p')\} 
\circ \eqclass{w}{\sim_{\ll}} \circ 
\bigcup_{(\overleftarrow{q\prime}\xrightarrow{c,-\gamma}(q,d))}  \{(q',q)\} \right ) 
$$
$$Z^{r}_{\rr,d} = 
\left \{ 
(p,q) \mid (p,\lmove,r,p',\rmove,\gamma) \in \delta^{\text{push}} \ , \ 
(p',\rmove,r,\gamma,q,d) \in
  \delta^{\text{pop}}
\right \}
$$
%
$$Z^{cw}_{\rr,d} = \bigcup_{\gamma \in \Gamma}  
\left ( \bigcup_{(\overleftarrow{p}\xrightarrow{r,+\gamma}\overleftarrow{p\prime})}  \{(p,p')\} 
\circ \eqclass{w}{\sim_{\rr}} \circ 
\bigcup_{(\overrightarrow{q\prime}\xrightarrow{r,-\gamma}(q,d))}  \{(q',q)\} \right ) 
$$
}

The expressions $Z^c_{\ll,d}$ and $Z^{cw}_{\ll,d}$ stands both for
left-to-left traversal reading twice the initial letter $c$; the
former one represents a back-and-forth move on $c$ whereas $Z^{cw}_{\ll,d}$ implies that between
the readings of $c$ a left-to-left traversal of $w$ is performed. If
the last direction $d$ is $\lmove$ then the reading head leaves the word, otherwise
the next reading will be $c$ again. The expressions $Z^r_{\rr,d}$ and
$Z^{wr}_{\rr,d}$ are defined dually. 

$$T^{cwr}_{\lr,d} = \bigcup_{\gamma \in \Gamma} 
\left ( \bigcup_{ (\overrightarrow{p}\xrightarrow{c,+\gamma}\overrightarrow{q}}  \{(p,q)\}  \right )
\circ  \eqclass{w}{\sim_{\lr}} \circ 
\left ( \bigcup_{ (\overrightarrow{p}\xrightarrow{r,-\gamma} (q,d))} \{(p,q)\}  \right )
$$
$$
T^{cwr}_{\rl,d} = \bigcup_{\gamma \in \Gamma} 
\left ( \bigcup_{(\overleftarrow{p}\xrightarrow{r,+\gamma}\overleftarrow{q}}  \{(p,q)\}  \right )
\circ  \eqclass{w}{\sim_{\rl}} \circ
\left ( \bigcup_{ (\overleftarrow{p}\xrightarrow{c,-\gamma} (q,d)}  \{(p,q)\}  \right ) 
$$

The expression $T^{cwr}_{\lr,d}$ represents a direct traversal from
left-to-right, going once through $c$ and $r$.

Finally, the classes $\eqclass{cwr}{\sim_{\ll}}$,
$\eqclass{cwr}{\sim_{\lr}}$,
$\eqclass{cwr}{\sim_{\rl}}$ and $\eqclass{cwr}{\sim_{\rr}}$ are
defined in Figure~\ref{fig:relations}.

\begin{figure*}[th]
$$\begin{array}{l}
\eqclass{cwr}{\sim_{\ll}}= 
\left ( Z^{c/cw}_{\ll,\rmove}  \cup \left (
  T^{cwr}_{\lr,\lmove} \circ (
Z^{r/wr}_{\rr,\lmove}  
)^* \circ T^{cwr}_{\rl,\rmove}
\right ) \right )^*  \left ( Z^{c/cw}_{\ll,\lmove} \cup \left (
  T^{cwr}_{\lr,\lmove} \circ (Z^{w/wr}_{\rr,\lmove})^* \circ T^{cwr}_{\rl,\lmove}
\right ) \right )\\

\eqclass{cwr}{\sim_{\lr}}= 
\left ( Z^{r/wr}_{\ll,\rmove} \cup
  \left ( T^{cwr}_{\lr,\lmove} \circ (Z^{r/wr}_{\rr,\lmove})^* \circ T^{cwr}_{\rl,\rmove}
\right )\right ) ^* \left ( T^{cwr}_{\lr,\rmove} \cup  \left (  T^{cwr}_{\lr,\lmove} \circ
  (Z^{r/wr}_{\rr,\lmove})^* \circ Z^{r/wr}_{\rr,\rmove}
\right ) \right )\\

\eqclass{cwr}{\sim_{\rl}}= 
\left ( Z^{r/wr}_{\rr,\lmove} \cup \left (
  T^{cwr}_{\rl,\rmove} \circ (Z^{c/cw}_{\ll,\rmove})^* \circ T^{cwr}_{\lr,\lmove}
\right ) \right )^*  \left ( T^{cwr}_{\rl,\lmove} \cup \left( T^{cwr}_{\rl,\rmove} \circ
  (Z^{c/cw}_{\ll,\rmove})^* \circ Z^{c/cw}_{\ll,\lmove}
\right ) \right )\\

\eqclass{cwr}{\sim_{\rr}}= 
\left ( Z^{c/cw}_{\rr,\lmove} \cup
  \left ( T^{cwr}_{\rl,\rmove} \circ (Z^{c/cw}_{\ll,\rmove})^* \circ T^{cwr}_{\lr,\lmove}
\right )\right ) ^*  \left ( Z^{r/wr}_{\rr,\rmove} \cup
  \left (T^{cwr}_{\rl,\rmove} \circ (Z^{c/cw}_{\ll,\rmove})^* \circ T^{cwr}_{\lr,\rmove}
\right ) \right )
\end{array}
$$
\caption{\label{fig:relations} Relations of the transition congruence
  on $cwr$,
where $Z^{c/cw}_{\ll,d}$ and  $Z^{r/wr}_{\ll,d}$ are defined
respectively by $Z^{c}_{\ll,d} \cup Z^{cw}_{\ll,d}$ and $Z^{r}_{\ll,d} \cup
Z^{wr}_{\ll,d}$.
}
\end{figure*}

Hence, we indeed have that $\eqclass{f_{c,r}(u)}{\sim_{\alpha\beta}} =
\eqclass{cur}{\sim_{\alpha\beta}} = \eqclass{cu'r}{\sim_{\alpha\beta}}
= \eqclass{f_{c,r}(u')}{\sim_{\alpha\beta}}$ for all $\alpha,\beta \in
\{\mathsf{l},\mathsf{r}\}$.
\qed
\end{proof}

\coremptinesstwovpa*
\begin{proof}
We detail the hardness proof by reduction from the emptiness of $k$ \dvpa.
The latter problem can be shown to be \exptime-hard from the \exptime-hardness of intersection emptiness of
$k$ deterministic top-down tree automata and a polynomial translation of deterministic top-down tree automata into \dvpa~\citeappendix{journals/jacm/AlurM09}. We can then
encode this problem as emptiness of a \dtwovpa as follows: the
\dtwovpa simulates one after the other the \dvpa's $A_1, \ldots, A_k$;
once the word is read from left to right simulating $A_i$, if a final
state of $A_i$ is reached, one enters a state that move the reading
head at the beginning of the word and then switches to the initial
state of $A_{i+1}$ to read the input word once again. Hence, starting
initially in the initial state of $A_1$, if the final state of $A_k$
is reached then the input nested word belongs to all the $A_i$'s.
\qed
\end{proof}

\subsection{Two-way visibly pushdown transducers}

\lemDecompUnVPT*
\begin{proof}
It has been proved in~\citeappendix{FS12} that every unambiguous \vpt can be transformed into
a \dvpt equipped with a look-ahead limited to the current hedge.
Formally, such a transducer is defined as a triple $(T,A,\lambda)$
where $T$ is a \dvpt, $A$ is a \vpa with no initial states, and 
$\lambda$ is a mapping from call transitions of $T$ to states of $A$.
Given a state $p$ of $A$, we denote by $A_p$ the \vpa defined
from $A$ by letting $\{p\}$ be the set of initial states.
A call transition $t$ of $T$ can then be fired at some position of an input word $w$ only 
if the longest nested subword of $w$ from this position belongs to $\cL(A_p)$.

Intuitively, the decomposition of a \dvpt with look-ahead works as follows: the
co-deterministic letter-to-letter \vpt does a first pass enriching the alphabet
with the results of the look-ahead tests.
Then the deterministic \vpt simulates the \dvpt with look-ahead using this additional information.

Formally, let $(T,A,\lambda)$ be a \dvpt with look-ahead from $\Sigma$ to $\Delta$, 
with $A=(Q,F,\Gamma,\delta)$.
We first define the structured alphabet $\Sigma'$ as the disjoint union of the set of call
symbols $\Sigma_c\times 2^Q$, and the set of return symbols
$\Sigma_r$.
We define the co-deterministic letter-to-letter \vpt $T_2=(A_2,\out_2)$ from $\Sigma$
to $\Sigma'$, where
$A_2$ is defined as the co-determinisation of $A$. Formally, let us denote by
$\id_X$ the set $\{(q,q) \mid q\in X\}$. We define
$A_2=(2^{Q\times Q},I_2, \id_F,\Sigma_r \times 2^{Q\times Q},\alpha)$ where 
$I_2=\{S\subseteq Q\times Q \mid S\cap I\times F \neq \emptyset\}$ 
and the transitions of $A_2$
are defined as follows:
\begin{itemize}
\item Return transitions: $\id_Q \xrightarrow{r, -(r,S)} S$, with $r\in \Sigma_r$ and $S\subseteq Q\times Q$
\item Call transitions: $S \xrightarrow{c, +(r,S')} S''$ with $c\in \Sigma_c$, $r\in \Sigma_r$ and
$S,S',S''\subseteq Q\times Q$, iff $S=\textsf{update}(c,S'',r)\circ S'$ where
$\textsf{update}(c,S'',r)=\{(p,q)\in Q\times Q\mid \exists \gamma \in \Gamma. \exists (p'',q'')\in S'' .
p\xrightarrow{c, +\gamma} p'' \text{ and }
q'' \xrightarrow{r, -\gamma} q \}$
\end{itemize}
\begin{figure}
\centering
\begin{tikzpicture}[scale=0.6]
\node (q1) at (2,-0.3) {$p''$};
\node (q2) at (5,-0.3) {$q''$};
\path[->] (q1) edge [above] node {$S''$} (q2);

\node (q) at (0,1) {$p$};
\node (qp) at (7,1) {$q$};

\path[->] (q) edge [right] node {$c,+\gamma$} (q1);
\path[->] (q2) edge [left] node {$r,-\gamma$} (qp);
\path[->,dashed] (q) edge [above] node {$Update(c,S,r)$} (qp);
\end{tikzpicture}
\end{figure}
The mapping $\out_2$ associates $\epsilon$ to every return transition, and associates 
$\{p\in Q \mid \exists q \in F. (p,q)\in S\}$ to the call transition $S \xrightarrow{c, +(r,S')} S''$. 
This set corresponds to the set of states $p$ such that the look-ahead constraint $p$ is satisfied.

The \dvpt $T_1$  from $\Sigma'$ to $\Delta$
can then easily be derived from the \dvpt with look-ahead $T$ as the look-aheads 
tests can be checked on the enriched alphabet $\Sigma'$.
\qed
\end{proof}

\ThmHUVPT*

\begin{proof}
We first notice that since we're considering visibly pushdown machines, the stacks of both machines are always synchronized, meaning that they have the same height on each position.
Then, let us remark that when the $\twovpt$ moves to the right, 
we can do the simulation in a straight forward fashion by simulating it on the production of the one-way, which we can compute.
It becomes more involved when it moves to the left.
We then need to rewind the run of the one-way, and nondeterminism can arise.
To bypass this, let us recall that a similar construction from~\citeappendix{HopUll67} exists for classical transducers, and that the rewinding is done through a back and forth reading of the input, backtracking the run up to a position where the nondeterminism is cleared, and then moving back to the current position.

The main idea is that if we were to consider a hedge as a word over subhedges (see Figure~\ref{Figure:WordSubhedgeAp}), we can use the Hopcroft-Ullman construction, given that we know the initial state, i.d. the state in which the one-way enters the hedge.
To overcome this, we will ensure the invariant that the stack contains not only the stack symbols from the two transducers, but also at each step it contains the state in which the one-way enters  a hedge.
Remark that thanks to this, upon moving to the left of a call letter, the state of the one-way is directly given by the information in the stack.

We now explain how we can treat subhedges as letters.
First, while the subhedge alphabet is infinite, we are actually interested in their behaviour in the one-way.
Thus we consider an automaton not over the subhedge, but over their summaries, which are finite.
We can thus compute a finite automaton of the summaries, and apply the Hopcroft-Ullman construction on it.
Consequently, we need to be able to compute the summaries of a given subhedge. This is easily done on the fly using the determinisation procedure of the $\vpa$s.
Finally, note that applying the Hopcroft-Ullman construction to the automaton of summaries gives the state in which the one-way enters the previous subhedge (when rewinding a run).
This allows us to maintain the invariant, and by reading this subhedge we can compute the state of the one-way at the previous position (from where we started).

Note that the Hopcroft-Ullman routine is deterministic, and consequently the construction preserves determinism.

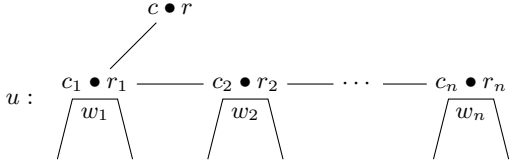
\begin{figure}
\begin{center}

\begin{tikzpicture}

\node (cr) at (5,5) {$c\bullet r$};

\foreach \i/\n in { 1/1, 3/2, 6/n}
{
\node (cr\n) at (3+\i,4) {$c_\n\bullet r_\n$};

\node (w\n) at (3+\i,3.6) {$w_\n$};

\draw (2.5+\i,3) -- (2.7+\i,3.8) -- (3.3+\i, 3.8) -- (3.5+\i,3);
}

\node (dots) at (7.5,4) {$\cdots$};
\node (u) at (3,3.8) {$u:$};
\draw (cr) -- (cr1) -- (cr2) -- (dots) -- (crn);

\end{tikzpicture}
\caption{The well nested word $cc_1w_1r_1c_2w_2r_2\ldots c_nw_nr_n$ can be seen as a word $u$ over letters $(c_i,S_i,r_i)$ where $S_i$ is the summary of $w_i$. The position labelled by $c$ serves as initial position of the word. }\label{Figure:WordSubhedgeAp}

\end{center}
\end{figure}

\paragraph{Formal construction.}
Let $A=((Q,i,F,\Gamma,\delta),\out_1)$ be a letter-to-letter $\dvpt$ and $B=((P,j,G,\Theta,\alpha),\out_2)$ be a $\twovpt$ that can be composed with $A$.
We assume that $A$ works on the alphabet equipped with left and right markers and preserves them. Note that it can easily be extended if it is not the case.

We construct $C=((N,k,H,\Omega,\beta),\out_3)$ a $\twovpt$ that realizes the composition.

\begin{itemize}
\item $N=N_m \uplus N_b \uplus N_f \uplus N_s$ where
$N_m$, $N_b$ and $N_f$ correspond to the classical sets of the Hopcroft-Ullman construction, and $N_s$ is used to compute the summary of a subhedge. 
We have the main mode $N_m=P\times Q$, the back mode $N_b=P\times Q \uplus P\times Q^Q$ and the further mode $N_f=P\times Q^2$, while $N_s=Q^Q$.
Note that there are also other states like $read$ or states from $P\times Q \times \{end\}$ that were omitted. The total size of the omitted states is linear in $P$ and $Q$.

\item $k=(i,j)$ is the initial state.
\item $H=F\times G$ is the set of final state.

\item $\Omega$ can similarly to $N$ be written as the disjoint union of stack alphabets for the different modes.
We have $\Omega_m=Q\times \Gamma \times \Theta$,
$\Omega_b = P\times (Q^Q\uplus Q)\times \Sigma_r$,
$\Omega_f= P \times ((Q\times \Gamma) \uplus (Q\times \Gamma)^2)$ and 
$\Omega_s=Q^Q\times \Sigma_r$.
\end{itemize}

We now give the transition function $\beta$.
Lowercase letters denote element of its uppercase counterpart. 
The direction of a transition is given by the sense of an arrow, and the resulting direction is omitted if it doesn't change. Push transitions are denoted with a $+$ symbol while pop transitions are denoted by a $-$ symbol.
For example, we write $(q,\leftarrow,r,q',\leftarrow,\gamma)$ in $\delta^{push}$ as
$q'\xleftarrow{r, +\gamma} q$ and $(q,\rightarrow,r, q',\leftarrow,\gamma)$ in $\delta^{pop}$ as
$q\xrightarrow{r,-\gamma} q',\leftarrow$.

\begin{itemize}
\item The first three items describe the cases when we are able to directly advance in the two runs. These are the simpler cases.
The fourth corresponds to the end of the Hopcroft-Ullman construction, where all the needed information was computed.
In these cases, the production of $\out_3$ is the one of the corresponding transition of $\out_2$. Note that in all other cases, the production of $\out_3$ will be empty and thus omitted.
\begin{itemize}

\item $(p,q) \xrightarrow{c,+(q,\gamma,\theta)} (p',q'),d$ if $\out_1(q\xrightarrow{c,+\gamma} q')=c'$ and $p\xrightarrow{c',+\theta}p',d$. 

\item $(p,q) \xrightarrow{r,-(q,\gamma,\theta)} (p',q'),d$ if $\out_1(q\xrightarrow{r,-\gamma} q')=r'$ and $p\xrightarrow{r',-\theta}p',d$.

\item $(p',q'),d\xleftarrow{c, -(q',\gamma,\theta)} (p,q)$
	if $\out_1(q'\xrightarrow{c,+\gamma} q)=c'$ and $p',d\xleftarrow{c',-\theta} p$.
	
\item $(p',q'),d \xleftarrow{r , +(q,\gamma,\theta) } (p,q,q',\gamma)$ where there exists $q''$ such that 
$\out_1(q'\xrightarrow{r,-\gamma} q'')=r'$ and
$p',d\xleftarrow{r',+\theta} p$.

\end{itemize}

\item When $B$ moves to the left on a recall letter, 
we engage in the Hopcroft-Ullman construction.
In order to do that, we need to compute the summary of the subhedge we are about to read.
Note that a similar transition happens when the automaton on summaries rewinds one more step.
Thus we have the following transitions:
\begin{itemize}
\item $\id_Q \xleftarrow{r, +(p,q,r)} (p,q)$.
\item $\id_Q \xleftarrow{r,+(p,R,r)} (p,R)$.
\end{itemize}

\item Computing a summary amounts to determining a $\vpa$. Note that we stop when we reach the height we are interested in, which is where the stack first contains a state of $B$, which is handle by the next item.
Given a summary $S$ and $c,r$ a call and return letter respectively, we define $Update(c,S,r)=\{(q,q')\mid \exists (q_1,q_2) \in S \text{ and } \gamma \ q\xrightarrow{c,+\gamma} q_1 \text{ and }
 q_2\xrightarrow{r,-\gamma} q' \}$. This will reveal to be useful in the remainder of the construction.
\begin{itemize}
\item $\id_Q \xleftarrow{r,+(S,r)} S$.
\item $S'' \xleftarrow{c,-(S',r)} S$ where 
 $S''=S'\circ Update(c,S,r)$. 
\end{itemize}

\item After reading the first subhedge, we get to the point where the top stack symbol is  of the form $(p,q,r)$. If there is only one candidate, then there is no ambiguity. Otherwise, we start rewinding the runs.
\begin{itemize}
 \item $ (p,q',end), \rightarrow\ \xleftarrow{c,-(p,q,r)} S$ if $q'$ is the only state such that $(q',q)$ belongs to $Update(c,S,r)$.
 
\item $(p,R) \xleftarrow{c,-(p,q,r)} S $ where 
$R=\{(q',q') \mid (q',q)\in Update(c,S,r)\}$.
\end{itemize}

\item After reading the following subhedges, similar subcases appear, depending on whether the nondeterminism is cleared or not.
If there is only one candidate left, we store a state leading to from the next subhedge, as well as a state leading to another candidate. 
They will be used to know when we got to the correct position.
Otherwise we just update the set of partial runs.
\begin{itemize}
 \item $ (p,q,q'),\rightarrow\ \xleftarrow{c,-(p,R,r)} S$ if 
 $R(q)$ is defined, $R\circ Update(c,S,r) \subseteq Q\times\{R(q)\}$
 and if $R(q')$ is defined and different from $R(q)$.
 
\item $(p,R') \xleftarrow{c,-(p,R,r)} S $ where 
$R'=R\circ Update(c,S,r)$ and $R'\not\subseteq Q\times \{q\}$ for any $q$.
\end{itemize}

\item It can happen that the nondeterminism has not been cleared until we reach the beginning of the hedge.
In the same way that the Hopcroft-Ullman uses the initial state, we then use the information on the top of the stack to decide the candidate.
\begin{itemize}
\item $(p,q,\theta,q') ,\rightarrow\ \xleftarrow{c,-(q,\gamma,\theta)}(p,R)$ where if $q''$ is such that $q\xrightarrow{c,+\gamma} q''$, both $R(q')$ and $R(q'')$ are defined and different. 
\end{itemize}

\item Due to the definition, the model of $\twovpt$ does not allow for direct u-turns. Consequently, the u-turns have been parametrized by specific states in the previous cases. We know explicit how we handle them:
\begin{itemize}
\item $(p,q,end )\xrightarrow{c,+(p,q,\gamma)} q'$ where 
 $q\xrightarrow{c,+\gamma} q'$. We also have a subroutine that follows run of $A$ on this subhedge until it ends.
\item $(p,q,q') \xrightarrow{c,+(p,q,q')} read$ where the state $read$ is a subroutine that only reads the subhedge until it pops the stacked information. 
\item $read \xrightarrow{r,-(p,q,q')} (p,q,q')$.
At the end of the $read$ subroutine, we start following two runs in parallel, in the same way as in the next subcase.
\item $(p,q,\theta,q') \xrightarrow{c,+(q,\gamma,\theta)} (p,q'',q')$ where $q\xrightarrow{c,+ \gamma} q''$.
\end{itemize}

\item When we are in states of $N_f$, i.d. states of the form $(p,q,q')$, we simply follow the two runs in parallel, stacking $p$ and the current states on the current height and the stack letters of both runs at each step nonetheless.
This subroutine ends upon popping a stack letter that contains $p$ where the two runs collide, meaning we reached the original position. We now explicit what happens on this position:
\begin{itemize}
\item $q\xrightarrow{ r,-(p,q',\gamma)} (p,q',q,\gamma),\leftarrow$.
\item $(q,q') \xrightarrow{ r,-(p,q_1,\gamma,q_2,\gamma')} (p,q_1,q,\gamma),\leftarrow$ if there exists $q''$ such that 
$q\xrightarrow{r,-\gamma} q''$ and 
$q'\xrightarrow{r,-\gamma'} q''$.
\end{itemize}
\end{itemize}
\qed

\end{proof}

\subsection{Expressiveness of Two-Way Visibly Pushdown Transducers}

\ThmorderPreserv*

\begin{proof}
The proof relies on the similar result for finite words from~\citeappendix{conf/icla/Filiot15} and 
the equivalence between \vpa and \MSONW from~\citeappendix{AM04}.
Let $T$ be a functional $\VPT$. From~\citeappendix{FS12}, we know that we can construct an equivalent unambiguous $\VPT$ $T'$ realizing the same function.
Using~\citeappendix{AM04}, we can construct an \MSONW formula $\varphi$ of the form $\exists X_1\ldots \exists X_{|Q|}\  \psi(X_1,\ldots,X_{|Q|})$ that recognizes $dom(T')$. Moreover, given $u$ in $dom(T')$, there exists a unique assignment of the variables $X_i$ satisfying $\psi$, such that a variable $x\in X_i$ if, and only if, $x$ quantify a position $j$
such that the unique accepting run of $T'$ on $u$ is in state $q_i$ on position $j$.
Using $\varphi$, we can then easily construct an \MSONWtoW transduction $T''$ using $|Q|$ copies.
The domain formula is $\varphi$, position formulas are $\phi_{pos}^q(x)=\varphi\wedge x\in X_q$.
The successor transition is given by $\phi_S^{q,q'}(x,y)=S(x,y)\wedge \phi_{pos}^q(x)\wedge \phi_{pos}^{q'}(y)$ and we label the $q$ copy of a node by the possibly empty production of the transducer in state $q$ reading the label of the node. We have, for $v$ a production of $T$, $\phi_v^q(x)=\bigvee_{a\in A_{q,v}} a(x)$ where $A_{q,v}=\{a\in A\mid \exists q'\ q\xrightarrow{a\mid v} q'\}$. 
Note that labeling by possibly empty words is not restrictive as MSO transductions are closed under composition, and a simple transduction can extend words into linear graphs and compress the $\epsilon$-labeled paths.

Now given an order-preserving \MSONWtoW $T$, we construct an unambiguous \VPT that recognizes the same function.
As $T$ is order-preserving, for every $u=u_1\ldots u_n$ in $dom(T)$, we can decompose $T(u)$ in $v_1\ldots v_n$ where $v_i$ corresponds to the production from position $i$.
Let us call $B$ the finite set of all possible $v_i$ appearing in a such decomposition.
For any $v$ in $B$, we use the formulas of $T$ to construct a formula 
$\phi_v(x)$ that holds on an input word $u$ and a position $i$ if in the decomposition of $T(u)$, $v_i=v$.
For any sequence $I=(c_1,\ldots,c_k)$ of $|v|$ different copies of $T$, we define
$\phi_I^v(x)= \bigwedge_{i<k} \phi_S^{c_i,c_{i+1}}(x,x)\wedge \bigwedge_{i\leq k} \phi_{pos}^{c_i}(x)\wedge \phi_{v_i}^{c_i}(x)\wedge \bigwedge_{x\notin I}\lnot \phi_{pos}^c(x)$.
The formula $\phi_v(x)$ is simply defined as the disjunction of the formulas $\phi_I^v(x)$ on all possible sequences $I$. 

Then using these formulas, we construct a formula $\psi$ over the finite alphabet $\Sigma\times B$
that recognizes the language $\cL=\{(u,T(u))\mid u\in dom(T)\}$.
We define $\psi= \phi'_{dom} \wedge \forall x\ (a,v)(x) \to \phi_v(x)$ where $\phi'_{dom}$ is obtained from $\phi_{dom}$ by replacing every predicate $a(x)$ by $\bigvee_{v\in B} (a,v)(x)$.
Now thanks to~\citeappendix{AM04}, we can construct a \dvpa that recognizes $\cL=\cL(\psi)$.
Finally, we transform it into a \VPT by replacing transitions reading $(a,v)$ into transitions reading $a$ and producing $v$.
Since $T$ realizes a function, we obtain a functional $\VPT$, concluding the proof.\qed
\end{proof}

\PropDecidSU*

\begin{proof}
We prove that this problem is equivalent to deciding the emptiness of a $\dtwovpa$, which concludes the proof thanks to Corollary~\ref{Cor-Emptiness2VPA}.

Let us first remark that if $A$ is single use, it is single use with respect to the set of all states that can produce a non empty word.
Let $A$ be a $\twovpt$ on an input alphabet $\Sigma$.
We define a $\twovpa$ $B$ on the marked alphabet $\Sigma\times \{0,1\}$ as follows.
The transducer $B$ first reads its input to ensure that there is exactly one position with a $1$.
It then nondeterministically chooses a producing state $q$ and simulates $A$ on its input.
It finally accepts if it visits the marked position twice in state $q$.
Then $A$ is single use if, and only if, the language recognised by $B$ is empty.
Since the size of $B$ is linear in the size of $A$, deciding the single use property is Exptime.

Conversely, let $B$ be a $\twovpa$.
We construct a $\twovpt$ $A$ as follows.
All existing transitions of $B$ are set to produce the empty word, and every accepting transition
is replaced by a back and forth move on the last position, producing a single letter.
Then the producing transitions can only be fired in $A$ if there is a run of $B$ that fires an accepting transition.
If it is the case, then the corresponding run on $A$ will visit the state $q$ twice in the last position while producing non empty words.
Thus the language recognised by $B$ is empty if, and only if, $A$ is single-use.
As the size of $A$ is linear in the size of $B$, we get the Exptime-hardness of the single use problem.
\qed
\end{proof}

\lemsula*
\begin{proof}
We explicit a transformation that is definable by an \MSONWtoW transduction but not by a strongly single-use \dtwovptla.

Consider an alphabet $\Sigma$ with some special letters $c$ and $r$ from $\Sigma_c$ and $\Sigma_r$ respectively.
We define the transformation $f$ which associates to a word $w_0c w_1 c w_2 \ldots w_{n-1} c w_n r w'_{n-1} \ldots w'_2 rw'_1 r w_0$ where all $w_i$, $w'_i$ are non empty nested words and do not any contain $c$, for $0\leq i\leq n$, the word $w_0 w'_0 w_1 w'_1\ldots w_{n-1}w'_{n-1} w_n$.
Its domain is then the set of nested words where any $c$ is matched by an $r$, and all letters $c$ appear successively nested on a given branch. The transformation is illustrated in Figure~\ref{Figure-sula}.

\begin{figure}
\begin{tikzpicture}[scale=0.5]

\foreach \i/\n/\j in { 0/0/4, 1/2/3, n-1/6/1}
{
\node (w\i) at (\n,\j) {$w_{\i}$};
\node (wp\i) at (16-\n,\j) {$w'_{\i}$};
}

\foreach \i/\j in { 1/4, 3/3, 5/2,7/1}
{
\node (c\i) at (\i,\j) {$c$};
\node (r\i) at (16-\i,\j) {$r$};
}

\node (dots1) at (4,2) {$\cdots$};
\node (dots2) at (12,2) {$\cdots$};
\node (wn) at (8,0) {$w_n$};

\path[->] (w0) edge [bend left=20] (wp0);
\path[->] (wp0) edge [bend right=20] (w1);
\path[->] (w1) edge [bend left=20] (wp1);
\path[->,dashed] (wp1) edge [bend right=20] (dots1);
\path[->,dashed] (dots1) edge [bend left=20] (dots2);
\path[->,dashed] (dots2) edge [bend right=20] (wn-1);
\path[->] (wn-1) edge [bend left=20] (wpn-1);
\path[->] (wpn-1) edge [bend right=20] (wn);

\end{tikzpicture}
\caption{The transformation $f$ alterns $n$ times between positions left and right of $w_n$. Thus it has to read $w_n$ at least $2n$ times.}\label{Figure-sula}
\end{figure}
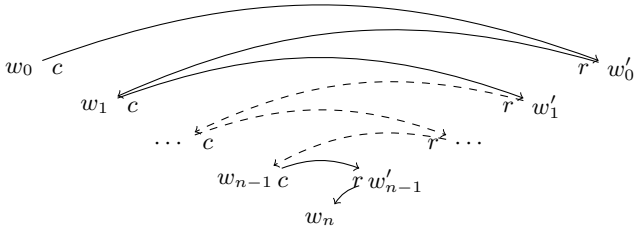

Before giving the \MSONWtoW that defines $f$, we explain how it is not definable by a strongly single use \dtwovptla.
As the $w_i$ and $w'_i$ are unbounded, they cannot be guessed by look-around. 
Thus a machine realizing it has to visit these subwords in the order they are output.
But each walk from $w_i$ to $w'_i$ has to cross $w_n$.
Thus $w_n$ is read at least $2n$ times. As $n$ is not bounded, $f$ cannot be realized by a strongly single use \dtwovptla.

Now we define a \MSONWtoW $T$ that realizes $f$.
In order to do that, we define a binary predicate $H(x,y)$ which holds is $x$ and $y$ are call or return positions of a same hedge.
Let $H_{tc}(X)$ be defined by the formula:
\begin{align*}
\forall x\in X\ & 
				\Sigma_c(x)\to \big(\forall y\ M(x,y) \vee (S(y,x)\wedge \Sigma_r(y))\to y\in X \big) \\
			\wedge\	&	
				\Sigma_r(x)\to \big(\forall y\ M(y,x) \vee (S(x,y)\wedge \Sigma_c(y))\to y\in X \big)		
\end{align*}

with $\Sigma_l(x)=\bigvee_{\sigma\in \Sigma_l} \sigma(x)$ for $l=c,r$.
Then a set $X$ satisfies $H_{tc}(X)$ if, and only if, it is closed by the relation \emph{belong to the same hedge}.
We then simply set $H(x,y)=\forall X\ x\in X \wedge H_{tc}(X)\to y\in X$.
We also define the parent relation 
$P(x,y)=\exists z\ H(x,z)\wedge (\forall z'\ H(z,z')\to z\leq z')\wedge S(y,z)$ which holds if $y$ is the call corresponding to the parent of $x$.

We can now define the domain formula
$\phi_{dom}=\forall x\ c(x) \to (\forall y \ (M(x,y)\to r(y))\wedge (P(x,y)\to c(y)) \wedge (H(x,y)\wedge c(y) \to x=y)$ stating exactly what was mentioned earlier.
The transducer $T$ uses $1$ copy, the position formula
$\phi_{pos}(x)= \lnot(c(x)\vee (\exists y\ M(y,x)\wedge c(y)))$ simply erases the $c$ labeled positions and their matching, 
the labeling formulas simply maintain the labels, and finally the successor formula $\phi_S(x,y)$ is defined by:
\begin{align*}
&\exists z\ S(x,z)\ \wedge  \lnot(c(z)\vee (\exists w\ M(w,z)\wedge c(w))) \wedge y=z)\\ 
 		 &\vee\ \big(  c(z) \wedge \exists z'\ M(z,z') \wedge S(z',y)\big)\\ 
		 &\vee\ \big( \exists z',z'' \ M(z',z)\wedge c(z')\wedge Next_c(z',z'') \wedge S(z'',y)\big) 
\end{align*}
where $Next_c(x,y)= x<y \wedge c(y) \wedge \forall z\ x<z<y \to \lnot c(z)$.
\qed
\end{proof}

\thmMSOtDVPTLA*

\begin{proof} We prove the equivalence $\MSONWtoW = \dtwovptlasu$.

    \textit{Proof overview} This result is based on several results from \citeappendix{journals/eatcs/CourcelleE12}, on the class of 
    deterministic tree-to-word walking transducers (DTWT), possibly
    augmented with visibly pushdown stack (then denoted DPTWT) and a regular look-around
    ability (denoted by an exponent ${la}$), and possibly restricted
    to linear-size increase the class of linear-size increase
    transductions (denoted by subscript $lsi$), or to strongly
    single-use (denoted by subscript $ssu$). We will define the most
    general model formally in the sequel.

    Let us also denote by \MSOTtoW the class of MSO-definable
    transductions from (ranked) trees to words. 
    Then, it is shown in \citeappendix{journals/eatcs/CourcelleE12}
    that 
    $$
    \MSOTtoW = \text{DTWT}^{la} = \text{DPTWT}^{la}_{lsi}
    $$
    The inclusion $\MSONWtoW \subseteq \dtwovptlasu$ is proved using the 
   equality $\MSOTtoW = \text{DTWT}^{la}$. Due to determinism, 
   $\text{DTWT}^{la}$ are always strongly single-use (otherwise they
   could be stuck in a loop), i.e., $\text{DTWT}^{la} =
   \text{DTWT}^{la}_{ssu}$ (see \citeappendix{journals/eatcs/CourcelleE12}, in
   which it is just called single-use). Using a first-child next-sibling encoding of nested
    words $w$ into binary trees $\fcns(w)$, we have 
    $\MSONWtoW = \MSOTtoW\circ \fcns$, and therefore 
    $\MSONWtoW = \text{DTWT}^{la}_{ssu}\circ \fcns$.     
    Then, we show that 
    $\text{DTWT}^{la}_{ssu}\circ \fcns\subseteq  \dtwovptlasu$ by simulating
    $\text{DTWT}^{la}_{ssu}$ that runs on \fcns encoding of nested words
    by $\dtwovptlasu$. In particular, when simulating tree walking
    moves, one do not preserve the strong single-use restriction, but
    the resulting \dtwovptlasu is single-use. 

    To show inclusion $\dtwovptlasu \subseteq \MSONWtoW$, we use
    the equality $\MSOTtoW = \text{DPTWT}^{la}_{lsi}$. Using \fcns encoding, we get that
    $\MSONWtoW
    = \text{DPTWT}^{la}_{lsi}\circ \fcns = (\text{DPTWT}^{la}\ \circ\
    \fcns)\cap\ \text{LSI}$, where LSI denotes the class of
    linear-size increase transductions. Then, we establish the
    inclusion $\dtwovptlasu \subseteq  (\text{DPTWT}^{la}\ \circ\ \fcns)\cap\
    LSI$. The single-use restriction of $\dtwovptlasu$ ensures that
    they define only transductions in LSI. Then,
    a $\dtwovptlasu$ can be simulated by a $\text{DPTWT}^{la}$ running on \fcns encodings of
    nested words. Due to the encoding, pushdown moves of the $\dtwovpt$
    are simulated by pushdown moves to the 1st child by the $DPTWT$
    and the look-around are translated in a straightforward fashion.
    The resulting $\text{DPTWT}$ does not use the stack while moving
    to 2nd children.

   As a matter of fact, putting things altogether, our result also shows
   that $\text{DPTWT}^{la}$ could be strengthen when they run on binary
   trees, to the following pushdown behaviour, while retaining
   MSO-expressiveness: they only need to push a symbol when moving to
   left children, and not when moving to right children. 

To summarize, we show the following chain of inclusions:
    {\small
    $$
    \begin{array}{c@{\hspace{5mm}}c@{\hspace{5mm}}c}
    \dtwovptlasu  & \subseteq\ (1) & (\text{DPTWT}^{la}\circ \fcns)\cap \text{LSI} \\ 
    \rotatebox{90}{$\subseteq$}\ (5)  & & \rotatebox{90}{$\supseteq$}\ (2) \\
    \text{DTWT}^{la}_{su}\circ \fcns & \supseteq\ (4) & \MSOTtoW \circ\fcns =
                                            \MSONWtoW\ (3)
    \end{array}
    $$
    }
    where (1) and (5) are shown in this paper, (3) is immediate, (2)
    and (4) come from \citeappendix{journals/eatcs/CourcelleE12}.
    We now proceed to the detailed proof. 

    \textit{Tree Walking Transducers}
    Let $\Lambda$ be a ranked
    alphabet of binary and constant symbols (i.e.$\Lambda$ is
    partitioned into $\Lambda_2$ and $\Lambda_0$). A tree $t$ of
    $\Lambda$ is a term inductively defined by $t ::= f(t,t)\ |\ a$,
    where $f\in\Lambda_2$ and $a\in\Lambda_0$. We denote by 
    $\text{Trees}_\Lambda$ the set of trees over $\Lambda$. The set of
    nodes, denoted by $N_t$ of a tree $t\in\text{Trees}_\Lambda$, is a
    prefix-closed subset of $\{1,2\}^*$ inductively defined as 
    $N_a = \{ \epsilon\}$ and $N_{f(t_1,t_2)} = \{ \epsilon\}\cup \{
    i.\pi\mid \pi\in N_{t_i}, i\in\{1,2\}\}$. For a node $n\in N_t$, 
    we denote by $t(n)$ the label of node $n$ in $t$. Let $\Sigma$ be a
    finite (unranked) alphabet. A tree to word transduction is a
    function from $\text{Trees}_\Lambda$ into $\Sigma^*$.

    Let us explain informally the different classes of tree walking
    transducers we consider in this proof. A deterministic tree to word walking
    transducer walks through the edges of a binary tree (starting from
    the root node), and
    writes a word from left to right on some output tape. In a state
    $q$ of a tree $t$ and at a node $n$, depending on the label of
    $n$, and the state $q$, the transducer can move either to 
    the father of $n$ (if it exists, otherwise the run rejects), the
    first or second child of $n$ (if it exists, otherwise the run
    rejects), change its internal state to a new state, and produces
    some partial word on the output. It can also decide to stop the
    walk by going to a stopping state $q_s$.

    Such transducers can be augmented with look-around. We define
    look-around by an unambiguous bottom-up tree automaton. Prior to 
    starting the computation of the tree walking transducer, the tree,
    if accepted by the look-around automaton, is labeled by the states
    of the accepting run of the automaton. Then, transitions are taken
    depending also on the look-around states.

    Finally, walking transducers can be augmented by a (visibly)
    pushdown store. Initially at the root the pushdown stack contains 
    an initial symbol $\gamma_0$, and whenever the transducer goes
    one step downward, it has to push one symbol on the stack. If it
    moves one step upward, it has to pop one symbol. At any moment, it
    can also read the top symbol of the stack.

    Formally, a \emph{deterministic pushdown tree to word walking
      transducer with look-around}  from $\text{Trees}_\Lambda$ to
    $\Sigma^*$ is a tuple $T = (L, Q,q_0,q_s,\Gamma, \gamma_0, R)$ where $L$
    is an unambiguous bottom-up tree automaton\footnote{We refer the
      reader to \citeappendix{TATA07} for a definition of bottom-up tree automata} over a finite set of
    states $P$ (the look-around automaton), 
    $Q$ is a finite
    set of states, $q_0$ is the initial state, $q_s$ the stopping
    state, $\Gamma$ is a finite stack alphabet with initial symbol $\gamma_0$, $R$ is a transition function such that 
    $$
    R\ :\ Q\times P \times \Lambda \times \Gamma \rightarrow
    \Sigma^*\times (\{q_s\}\cup \Gamma\times Q \times \{1,2\}\cup Q\times \{-1\})
    $$
    A configuration of $T$ on a tree $t$ is a triple $(q,n,\beta,u)\in
    Q\times N_t\times \Gamma^+\times \Sigma^*$. For all trees
    $t\in\text{Trees}_\Lambda$, if $t$ is accepted by the look-around
    automaton, we define $\rightarrow_t$ a binary
    relation between consecutive configurations as follows:
    for all $q,q'\in Q$, all $n,n'\in N_t$, all $\beta,\beta'\in\Gamma^*$,
    all $\gamma\in\Gamma$, all $u,v\in\Sigma^*$, 
    $(q,n,\beta\gamma,u)\rightarrow_t (q',n',\beta',uv)$ if 
    the accepting run of $L$ labels $n$ by a state $p\in P$ such that 
    $(q,p, t(n),\gamma)\in \dom(R)$ and either 

    \begin{itemize}
        \item (\textit{stopping move}) $R(q,p,t(n),\gamma) = (v, q_s)$ and $q'=q_s$, $n'=n$,
          $\beta'=\beta\gamma$, or
        \item (\textit{downward move}) $R(q,p,t(n),\gamma) = (v, \gamma', q', i)$ for
          $i\in\{1,2\}$ and $\beta' = \beta\gamma\gamma'$,
          $t(n)\in\Lambda_2$, and $n'$ is the
          $i$-th child of $n$, or 

        \item (\textit{upward move}) $R(q,p,t(n),\gamma) = (v, q', -1)$ and $n\neq \epsilon$
          (i.e. $n$ is not the root node), $\beta' = \beta$, and $n'$ is the father of $n$. 
    \end{itemize}
    A run of $T$ on a tree $t$ is 
    a finite sequence of configurations $c_0c_1\dots c_m$ such that
    $c_i\rightarrow_t c_{i+1}$ for all $i=0,\dots,m-1$. It is
    accepting if $c_0 = (q_0, \epsilon, \gamma_0,\epsilon)$ and $c_m =
    (q_s, n, \beta, u)$ for some $n\in N_t$,
    $\beta\in\Gamma^+$, $u\in\Sigma^*$. 
    Since $R$ is a function and $L$ is unambiguous,
    there exists at most one accepting run per input tree $t$, and we
    call $u$ the output of $t$. The transduction realized by $T$ is 
    the set of pairs $(t,u)$ such that $t$ is accepted by the
    look-around automaton, and there exists an accepting run of $T$ on
    $t$ whose output is $u$. The class of deterministic pushdown tree
    to word walking transducers    \footnote{We have slightly changed the definition of
      \citeappendix{journals/eatcs/CourcelleE12} to simplify our presentation,
      but in an equivalent way, and have specialized it to the
      tree-to-word setting. In \citeappendix{journals/eatcs/CourcelleE12},
    look-around are MSO-formulas on trees, with one free first-order variable, attached to
    the transitions of the transducer: a transition can be fired only
    if its look-around formula holds at the current node. It is known that such an MSO
    formula $\phi(x)$ is equivalent to an unambiguous bottom-up tree
    automaton $A_\phi$ \citeappendix{conf/dbpl/NiehrenPTT05,NevSch02} in the following sense: the automaton as
    a special set of selecting states $S$, such that on a tree $t$
    accepted by the automaton, a node $n$ is such that $t\models
    \phi(n)$ iff this node is labeled by a state of $S$ in the
    accepting run of the automaton on $t$. If
    $\phi_1(x),\dots,\phi_n(x)$ are the look-around formulas 
    appearing on the transitions of the tree walking transducer, then 
    by taking the product of the unambiguous automata $A_{\phi_i}$,
    one obtains an unambiguous automaton $A_{la}$ such that, on a tree $t$
    accepted by $A_{la}$, the state label of a node $n$ of $t$ in the
    accepting run of $A_{la}$ contains enough information to decide
    which look-arounds $\phi_i(x)$ hold at that node or not. 

    Another modification of the definition of
    \citeappendix{journals/eatcs/CourcelleE12} is that we do not have
    $0$-moves, i.e. transitions that stay at the same node. This 
    is circumvented by adding the possibility of producing several
    symbols by a single transition, in contrast to
    \citeappendix{journals/eatcs/CourcelleE12}.

    Finally, our transducers produce words over $\Sigma^*$, while 
    in \citeappendix{journals/eatcs/CourcelleE12}, they produce unary
    trees, i.e. a sequence of unary symbols followed by a constant
    symbol. As a consequence, when a constant is produced, the
    transducers of \citeappendix{journals/eatcs/CourcelleE12} stop. In our
    definition, we rather have added a stopping state.
  } is denoted by $\text{DPTWT}^{la}$.

    The class $\text{DTWT}^{la}$ denotes the class of deterministic tree
    to word walking transducers with look-around (without pushdown
    store), defined similarly as $\text{DPTWT}^{la}$ but with a pushdown
    alphabet $\Gamma$ that consists of one symbol $\{\gamma_0\}$ only.  
    In that case, we can omit the pushdown symbols in the transitions,
    except the initial pushdown symbol $\gamma_0$ that allows to know
    whether the current node is the root of the tree or not. Instead
    of keeping the symbol $\gamma_0$ in the transitions, we use a
    boolean value which is true if the current node is the root, and
    false otherwise. Therefore, in $\text{DTWT}^{la}$, the transitions
    function has the following type:
    $$
    R\ :\ Q\times P \times \Lambda \rightarrow
    \Sigma^*\times (\{q_s\}\ \cup\ Q \times \{-1,1,2\})
    $$
    We say that a $\text{DTWT}^{la}$ is
    \emph{strongly single-use} if one any accepting run $c_0\dots c_n$ on a
    tree, a node $n\in N_t$ is not visited twice in the same state.
    As a matter of fact, it turns out that $\text{DTWT}^{la}$ are
    always strongly single-use \citeappendix{journals/eatcs/CourcelleE12}.

    $\MSOTtoW$ is defined similarly as $\MSONWtoW$, except that 
    MSO formulas over the signature $\{S_1(x,y),S_2(x,y),(a(x))_{a\in\Lambda}\}$ are
    used to define the transduction, where $S_i(x,y)$ holds if $y$ is
    the $i$-th child of $x$. It is easy to show that $\MSONWtoW$ and
    $\MSOTtoW$ are equivalent, modulo suitable encodings. In
    particular, we will need the following equality:
    $$
    \MSONWtoW = \MSOTtoW\circ \fcns\qquad (1)
    $$
    where $\fcns$ encodes nested words into binary trees. Let us
    describe this encoding formally: Let $\Sigma = \Sigma_c\uplus \Sigma_r$ be a structured alphabet. 
    We define the binary alphabet $\Lambda_2 = \Sigma_c\times
    \Sigma_r$, and the unary alphabet $\Lambda_0 = \{ \bot\}$. 
    We define $\fcns$ the encoding of nested words to binary trees
    inductively as follows, for all $w,w'\in\WN[\Sigma]$, $c\in\Sigma_c,r\in\Sigma_r$:
    $$
    \fcns(cwrw') = (c,r)(\fcns(w),\fcns(w'))\qquad \fcns(\epsilon) = \bot
    $$
    We are now equipped to show the following two inclusions, which
    will prove the desired result:
    $$
    (2)\ \ \text{DTWT}^{la} \circ \fcns \subseteq \dtwovptlasu \qquad $$
    $$(3)\ \ 
    \dtwovptlasu\subseteq (\text{DPTWT}^{la}\circ \fcns) \cap LSI
    $$

    \textit{Proof of inclusion (2)} Let $T = (L, Q,q_0,q_s, R)$ be
    $\text{DTWT}^{la}$. We construct a $\dtwovptlasu$ $T'$ such that
    $\inter{T}\circ \fcns = \inter{T'}$.

    Transitions of $T$ are simulated only on call symbols, i.e. the
    moves of $T$ are simulated by moves in $T'$ between call symbols.
    For instance, let $c_1c_2r_2c_3c_4r_4r_3c_5r_5r_1$ be a
    nested word, whose \fcns encoding is the tree
    $$(c_1,r_1)((c_2,r_2)(\bot, (c_3,r_3)((c_4,r_4),
    (c_5,r_5)(\bot,\bot))))$$
    If $T$ moves from $(c_3,r_3)$ to its father
    $(c_2,r_2)$, then $T'$ will move from $c_3$ to $c_2$. If $T$ moves
    from $(c_3,r_3)$ to its first child $(c_4,r_4)$, then $T'$ will
    move from $c_3$ to $c_4$. If $T$ moves from $(c_3,r_3)$ to its
    second-child $(c_5,r_5)$, then $T'$ will move from $c_3$ to
    $c_5$.

    Therefore, one-step moves of $T$ in a binary encoding $t_w$ of a
    nested word $w$ are simulated by sequences
    of moves of $T'$ in $w$. It is easy to see however that those sequences
    of moves can be achieved by a $\dtwovpt$. The trivial case is a
    first-child move: it suffices to move one-step to the right in
    $w$. For the second-child move, $T'$ has to move to the next call
    symbol at the right of current one, at the same nesting depth: this is
    done by pushing one special symbol $\gamma_2$ when reading the
    first call,  and moving to the right, until $\gamma_2$ is popped. For a father move, 
    it suffices to move left: if the previous symbol is a call, then
    $T'$ has arrived to the call corresponding to the father
    node. Otherwise, the left symbol is a return symbol: again, a special symbol
    $\gamma_f$ is pushed, to know, when $T'$ walks left, whenever
    it is at the same depth as the initial call symbol. If after
    popping $\gamma_f$, a call symbol is read again, then it corresponds to
    the father. Note that these walks do not produce anything on the
    output.

    The look-around $L$ of $T$ is transformed into a look-around of $T'$
    such that, if $L$ labels a tree node labelled $(c,r)$ by a state
    $p$, then $T'$ will label the call symbol $c$ by the state $p$, as
    well as the call symbol $r$. It is possible, since bottom-up tree
    automata and visibly pushdown automata correspond modulo
    first-child next-sibling encodings, while
    preserving unambiguity \citeappendix{journals/jacm/AlurM09}. Therefore, if $P$ is the set of
    states of $L$, then the set of states of the look-around automaton
    of $T'$ is $P\times \Sigma_r$.

    Then, a transition $(q, p, (c,r), u, (q',d))$ where
    $d\in\{-1,1,2\}$ is simulated by $T'$ by a sequence of transitions
    (depending on whether $d = -1$, $d=1$ or $d=2$)
    that starts in state $q$ (with look-around state $(p,r)$) and ends
    in state $q'$, and performs moves as explained before.

    There is a last additional technical difficulty: \fcns encodings contain the
    symbol $\bot$, unlike the encoded nested words. Therefore,
    $T'$ may move to $\bot$, while $T$ cannot. Moves to nodes labeled
    $\bot$ can be simulated easily by $T'$ by adding
    $\epsilon$-transitions, which can in turn be removed while
    preserving determinism. It is not difficult but unnecessarily technical. 

    Finally, since $T$ is necessarily single-use (due to
    non-determinism), $T'$ is also single-use (the extra states added
    to simulate one-step moves of $T$ by several moves of $T'$ may be
    used several times at the same tree node, but the transitions
    fired from those states are $\epsilon$-producing).

\textit{Proof of inclusion (3)} Due to the single-use restriction,
any \dtwovptlasu transduction is LSI. It remains to show that 
a \dtwovptlasu can be simulated by a $\text{DPTWT}^{la}$. By using again the
correspondence between (unambiguous) visibly pushdown automata and
(unambiguous) bottom-up tree automata, one can simulate their
look-arounds. Since $\text{DPTWT}^{la}$ have the ability to push stack
symbols in both directions (first-child or second-child), it is not
difficult to construct a $\text{DPTWT}^{la}$ that simulates a \dtwovptlasu. As a matter of fact, pushing
symbols when moving to the second-child is not necessary to simulate
\dtwovptlasu: indeed, a second-child in a \fcns encoding correspond to
a next-sibling in the nested word, and \dtwovptlasu do not use
their stack for processing symbols that are at the same depth (they do
not push ``horizontally''). \qed
\end{proof}

\subsection{Inclusion into streaming transducers and hedge-to-string transducers}

\EquivModels*

\begin{proof}
  The proofs of these two inclusions share a same intermediate formal
  description of transformation. It turns out that this representation
  will be an extension of the finite transition algebra
  $\algtra{A}$ for some \dtwovpa $A$.

  We recall that elements from the algebra $\algtra{A}$ are binary
  relations over $Q \times \Moves$ where $Q$ is the set of states of
  $A$ and can thus be depicted as Boolean square matrices
  $\mathcal{M}_{\algtra{A}}$ over $Q \times \Moves$. Hence, the
  morphism $\mu_{\algtra{A}}$ associates with each word $w$ from
    $\WN{\Sigma}$ a matrix from $\mathcal{M}_{\algtra{A}}$ such that 
$\mu_{\algtra{A}}(w)((p_1,d_1),(p_2,d_2))$ is true if there exists a run
on $w$ from $(p_1,d_1)$ to $(p_2,d_2)$ in $A$. 

  One may extend this notion to transducers as follows. For a \dtwovpt
  $A$, we consider square matrices $\mathcal{N}_A$ over $Q \times
  \Moves$ whose values range over subsets of $\Delta^*$. One can
  define a mapping $\mu$ from $\WN{\Sigma}$ to $\mathcal{N}_A$ such
  that for all words $w$ from $\WN{\Sigma}$, $\mu(w)$ is a matrix $N_A$
  satisfying that $N_A((p_1,d_1),(p_2,d_2))$ is equal to $L$ if for
  each $v$ in $L$, there exists a run on $w$ from $(p_1,d_1)$ to
  $(p_2,d_2)$ in $A$ producing $v$. Note in fact that $A$ being
  deterministic, $L$ is either a singleton or the empty set.
  One can actually prove that one can
  define an (infinite) algebra $\algtrans{A}=( \mathcal{N}_A,
  .^{\algtrans{A}}, {(f_{c,r})}_{(c,r) \in \Sigma_c \times \Sigma_r},
  \epsilon^{\algtrans{A}})$ such that $ .^{\algtrans{A}}$ is associative and 
  $\epsilon^{\algtrans{A}}$ is its neutral element. 
Moreover, the
  considered mapping $\mu$ turns out to $\mu^{\algtrans{A}}$ be the canonical morphism from 
$\algwn$ to $\algtrans{A}$. 
It is worth
  noticing that for all $(p,d)$, $(p',d')$, $\mu_{\algtra{A}}(w)((p,d),(p',d'))$ is false
  iff $\mu_{\algtrans{A}}(w)((p,d),(p',d')) \neq \emptyset$. 

 The operations  $\epsilon^{\algtrans{A}}$, $.^{\algtrans{A}}$ and
 $f_{c,r}^{\algtrans{A}}$ can be represented as matrices as well. 
 To do so, let us first consider the two sets of symbols $\Xi^\alpha=\{
  x^{(p,d),(p',d')}_\alpha \mid (p,d),(p',d') \in Q \times \Moves\}$
  for $\alpha \in \{1,2\}$. Then,
  let $\hat{\mathcal{N}}_A$ be the set of matrices defined over $Q
  \times \Moves$ such that for $\hat{N}_A$ in $\hat{\mathcal{N}}_A$, for all
  $(p_1,d_1)$, $(p_2,d_2)$, $\hat{N}_A((p_1,d_1),(p_2,d_2))$ is either the empty
  set $\varnothing$ or a singleton
  set included into the set of words $(\Delta \cup \Xi^1 \cup
  \Xi^2)^*$. Moreover, for $\epsilon^{\algtrans{A}}$, the matrix is
  precisely the one with $\{\epsilon\}$ on its main diagonal and
  $\varnothing$ everywhere else. For $f_{c,r}^{\algtrans{A}}$, the
  elements of the matrix are actually included into $(\Delta \cup
  \Xi^1)^*$. The operation $.^{\algtrans{A}}$ deals with two matrices
  $\hat{N}^1_A$ and $\hat{N}^2_A$ and produces the matrix $\hat{N}''_A$
  satisfying that for all $(p_1,d_1)$, $(p_2,d_2)$,
  $\hat{N}''_A((p_1,d_1),(p_2,d_2))$ 
 is obtained from $\hat{N}^{'.'}_A$  the matrix of  $.^{\algtrans{A}}$
 by replacing everywhere in
 $\hat{N}^{'.'}_A((p_1,d_1),(p_2,d_2))$ the symbol
 $x^{(p,d),(p',d')}_\alpha$ by $\hat{N}^\alpha_A((p,d),(p',d'))$
 for $\alpha \in \{1,2\}$. The application for
 $f_{c,r}^{\algtrans{A}}$ represented by some matrix
 $\hat{N}^{f_{c,r}}_A$ is similar 
 with a single matrix as operand. 

The matrices $\hat{N}^{'.'}_A$ and $\hat{N}^{f_{c,r}}_A$ can be defined by
means of expressions similar to the ones defining  recursively the equivalence
classes of traversals. Hence, these matrices are defined by means of
unions, concatenations and Kleene star entrywise; the fact that entries of the
matrices contain at most singletons and that Kleene star can be
expressed as finite concatenations relies on the determinism of $A$. 

 From the infinite algebra $\algtrans{A}$ and more specifically the
 matrices representation of the operators $\hat{N}^{'.'}_A$ and $\hat{N}^{f_{c,r}}_A$
 from this algebra, we are going now to build a 
 streaming tree-to-string transducer on the one side and a $\dHtoS$
 on this other side. 

 For streaming tree-to-string transducers, the idea is to define from
 a \dtwovpt $A$ such a machine $S_{A}$ to simulate the computation of $\mu^{\algtrans{A}}(w)$ for any word
 $w$ or more precisely to compute the value associated to $((q_I,\rmove),(q_f,\rmove))$ in this matrix,
 $q_I$ being the initial state
 of the \dtwovpt $A$ and $q_f$ the (final) state reached by $A$ after reading
 $w$ from $q_I$.  

We recall that we can define from $A$ the finite algebra $\algtra{A}$
whose domain is $\text{Trav}_A$ and we consider $\Xi^\alpha=\{
  x^{(p,d),(p',d')}_\alpha \mid (p,d),(p',d') \in Q \times \Moves\}$
  for $\alpha \in \{1,2\}$.  

For a \dtwovpa $A$, the \stst $S_{A}$ is defined by
$(\text{Trav}_A,\epsilon^{\algtra{A}}, \Sigma_c \times \text{Trav}_A, \Xi^1,\delta_{S_{A}},
\mu_F^{S_{A}})$ where 
 \begin{itemize}
\item
  $\delta^{push}_{S_{A}}(m^{\algtra{A}},c)=(\epsilon^{\algtra{A}}, (c,m^{\algtra{A}}),\nu_{Id})$
\item $\delta^{pop}_{S_{A}}=(m'^{\algtra{A}},r,
  (c,m^{\algtra{A}}))=(m^{\algtra{A}} .^{\algtra{A}}
  f^{\algtra{A}}_{c,r}(m'^{\algtra{A}}), \nu_{c,r})$
\end{itemize}
where $\nu_{Id}$ is the identity on $\Xi ^1$ and $\nu_{c,r}$ associate
with each $x^{(p,d),(p',d')}$ the expression from $N((p,d),(p',d'))$
where the matrix $N$ is defined as $N^1_{Id} .^{\algtrans{A}}
f^{\algtrans{A}}_{c,r}(N^2_{Id})$, the matrices
$N^\alpha_{Id}$ satisfying  for all
$(p_1,d_1)$, $(p_2,d_2)$,
$N^\alpha_{Id}((p_1,d_1),(p_2,d_2))=x^{(p_1,d_1),(p_2,d_2)}_\alpha$ (for all $\alpha \in \{1,2\}$). 

Let us now consider the case of deterministic hedge-to-string
transducer. We first define the bottom-up deterministic look-ahead
automaton $B_A$ as
$(\text{Trav}_A,\{\epsilon^\algtra{A}\},\delta_{B_A})$ where 
$\delta_{B_A}$ is the set of rules of the form $\{(m^{\algtra{A}} .^{\algtra{A}}
f^{\algtra{A}}_{c,r}(m'^{\algtra{A}}),c,r,m^{\algtra{A}},m'^{\algtra{A}})$
with $m^{\algtra{A}}, m'^{\algtra{A}} \in \text{Trav}_A$. 

Now, we define the \dHtoS $H_A$ as follows: the set of states
$Q_{H_A}$ is $\{q_I\} \cup \text{Trav}_A \times (Q \times \Moves)^2$,
$q_I$ is the initial state, and the set of final states is $Q_{H_A}$. 
Now, for the transition function, we define
\begin{align*}
& \delta(q_I,c,r,n_1^{\algtra{A}},n_2^{\algtra{A}}) = 
\\
& \qquad \omega\lbrack
x^{(q_1,d_1),(q_2,d_2)}_\alpha \leftarrow  (n_\alpha^{\algtra{A}},((q_1,d_1),(q_2,d_2)))(x_\alpha)
\rbrack
\\
 &\delta((m^{\algtra{A}},((p,d),(p',d'))),c,r,m^{\algtra{A}}) = \\  &
\qquad \omega\lbrack
x^{(q_1,d_1),(q_2,d_2)}_\alpha \leftarrow  (m_\alpha^{\algtra{A}},((q_1,d_1),(q_2,d_2)))(x_\alpha)
\rbrack
\end{align*}
where 
\begin{itemize}
\item for any $n_1^{\algtra{A}}$, $n_2^{\algtra{A}}$ such that
$(f_{c,r}^{\algtra{A}}(n_1^{\algtra{A}}) .^{\algtra{A}}
n_2^{\algtra{A}})((q_I,\rmove),(q_f,\lmove))$ is true for some $q_f \in
F$. 
\item for any $m_1^{\algtra{A}}$, $m_2^{\algtra{A}}$ such that
$f_{c,r}^{\algtra{A}}(m_1^{\algtra{A}}) .^{\algtra{A}}
m_2^{\algtra{A}} = m^{\algtra{A}}$ and $((p,d),(p',d')) \in
m^{\algtra{A}}$ and 
\item $\omega$ is equal to the word $N((p,d),(p',d'))$, $N$ being
the matrix $f^{\algtrans{A}}_{c,r}(N^1_{Id}) .^{\algtrans{A}}
N^2_{Id}$, where the matrix $N^{\alpha}_{Id}$ satisfies for all $\alpha
\in \{1,2\}$, for all $(p_1,d_1)$, $(p_2,d_2)$,
$N^{\alpha}_{Id}((p_1,d_1),(p_2,d_2))=x^{(p_1,d_1),(p_2,d_2)}_\alpha$.
\end{itemize}

Let us prove now that the inclusions are strict. The transformation
serving as a counter-example is the same for the two classes. 
We consider a transformation $T$ over the input alphabet $\Sigma_c=\{c\}$
  and $\Sigma_r=\{r\}$ and the output alphabet $\{a\}$. This
  transformation takes as an input words such as $(cr)^n$ for any
  natural $n$ and outputs $a^{2^n-1}$. $T$ is given by
  the $\dHtoS$ with a single state $q$ and with a universal look-ahead
  automaton with $q'$ as unique state by 
$$\delta(q,c,r,q')=aq(x_1)q(x_1)$$
The transformation $T$ can also be defined by a \stst with a single
state $q$, a unique register variable $X$ and a unique stack symbol
$\gamma$. The transitions are given by 
\begin{itemize}
\item $\delta^{push}(q,c)=(q,\gamma,\{X \mapsto aXX\})$
\item $\delta^{pop}(q,r,\gamma)=(q,\{X \mapsto X\})$ 
\end{itemize}

The transformation $T$ cannot be realized by some \dtwovpt; indeed,
for such a machine with stack alphabet $\Gamma$ on the shallow inputs
of the domain, the possible stacks occurring in runs are either $\bot$
or the form $\gamma$ for $\gamma \in \Gamma$. Hence, the possible
behaviours of such \dtwovpt are similar to the ones of a deterministic finite state
transducer. It is known that deterministic finite state transducers
realize only functions that are linear-size increase; this is not the case of the
transformation $T$. \qed 
\end{proof}

\subsection{Unranked Tree Walking Transducers}

\paragraph{Unranked Trees} Let $\Lambda$ be a finite set of symbols. 
\emph{Unranked trees} $t$ over $\Lambda$ are defined inductively as
$t ::= a\mid a(t_1,\dots,t_n)$, for all $a\in\Lambda$, all $n\geq 1$. 
Unranked trees over $\Lambda$ can be identified (modulo renaming of
nodes) with structures over
the signature $\mathbb{U}_\Lambda$ that consists of the
\emph{first-child} predicate $fc(x,y)$ that relates a node $x$ to its
first-child $y$, the \emph{next-sibling} predicate $ns(x,y)$ that
relates a node $x$ to its next-sibling $y$ in a sequence of unranked
trees, and $a(x)$, for all $a\in\Lambda$, that holds true in node $x$
if it is labeled $a$. In addition, we also add a \emph{parent} predicate
$parent(x,y)$ that relates a node to its parent.

For instance, the unranked tree $a(b,c(a),c)$ is identified with the
structure whose set of nodes is $\{\epsilon, 1, 2, 3, 21\}$, where the
first-child predicate is $\{ (\epsilon, 1), (2, 21)\}$, the
next-sibling predicate is $\{(1,2),(2,3)\}$, the $a$ predicate is 
$\{\epsilon, 21\}$, the $b$ predicate is $\{1\}$ and the $c$ predicate
is $\{2,3\}$. The parent predicate is given by
$\{(1,\epsilon),(2,\epsilon),(3,\epsilon),(21,2)\}$.

\paragraph{Unranked Tree Walking Transducers} They are defined
similarly as ranked tree walking transducers, except that they move
along the next-sibling and first-child predicates. They are equipped
with a (visibly) pushdown store such that whenever they go down 
the first-child, they have to push some symbol, whenever they go up
to the parent of a node, they have pop one symbol from the
stack. However, when they move horizontally along next-sibling
predicates, they do not touch the stack. Before applying a transition,
they can test whether the current node is the root, is the first-child
of some node, the last-child, or a leaf. Their move have to be
consistent with the result of such a test. They are also equipped with
\emph{stay} moves that stay at the same tree node.

Formally, a \emph{deterministic pushdown unranked tree to word walking
      transducer} (DPT$_u$WT) from unranked trees over $\Lambda$
    to $\Sigma^*$ is a tuple $T = (Q,q_0,q_s,\Gamma, \gamma_0, R)$ where 
    $Q$ is a finite
    set of states, $q_0$ is the initial state, $q_s$ the stopping
    state, $\Gamma$ is a finite stack alphabet with initial symbol $\gamma_0$, $R$ is a transition function such that 
    $$
    R\ :\ Q\times \Lambda \times \Gamma \times \{0,1\}^4\rightarrow
    \Sigma^*\times (\{q_s\}\cup \Gamma\times Q \times \{\downarrow\}\cup Q\times \{\rightarrow,\leftarrow,\uparrow,\circlearrowleft\})
    $$
    A configuration of $T$ on a tree $t$ with set of nodes $N_t$ is a triple $(q,n,\beta,u)\in
    Q\times N_t\times \Gamma^+\times \Sigma^*$. We define $\rightarrow_t$ a binary
    relation between consecutive configurations as follows: Let $n\in
    N_t$ labeled $a\in \Lambda$. Let $\overline{b} = (b_{fc},b_{lc},b_r,b_l)\in
    \{0,1\}^4$ such that $b_{fc}=1$ iff $n$ is a first-child,
    $b_{lc}=1$ iff $n$ is a last-child, $b_r=1$ iff $n$ is the root,
    $b_l=1$ iff $n$ is a leaf. Then, for all $q,q'\in Q$, all $n'\in N_t$, all $\beta,\beta'\in\Gamma^*$,
    all $\gamma\in\Gamma$, all $u,v\in\Sigma^*$, 
    $(q,n,\beta\gamma,u)\rightarrow_t (q',n',\beta',uv)$ if 
    \begin{itemize}
        \item (\textit{stopping move}) $R(q,a,\gamma,\overline{b}) = (v, q_s)$ and $q'=q_s$, $n'=n$,
          $\beta'=\beta\gamma$, or
        \item (\textit{downward move}) $R(q,a,\gamma,\overline{b})
          = (v, \gamma', q', \downarrow)$, $\beta' =
          \beta\gamma\gamma'$, and $fc(n,n')$, or,

        \item (\textit{upward move}) $R(q,a,\gamma,\overline{b}) = (v,
          q', \uparrow)$ and $\beta' = \beta$, and $parent(n,n')$, or,

        \item (\textit{left sibling move}) $R(q,a,\gamma,\overline{b}) = (v,
          q', \leftarrow)$ and $\beta' = \beta$, and $ns(n',n)$, or,

        \item (\textit{right sibling move}) $R(q,a,\gamma,\overline{b}) = (v,
          q', \rightarrow)$ and $\beta' = \beta$, and $ns(n,n')$.

        \item (\textit{stay move}) $R(q,a,\gamma,\overline{b}) = (v,
          q', \circlearrowleft)$ and $\beta' = \beta$, and $n=n'$.

    \end{itemize}

    A run of $T$ on an unranked tree $t$ is 
    a finite sequence of configurations $c_0c_1\dots c_m$ such that
    $c_i\rightarrow_t c_{i+1}$ for all $i=0,\dots,m-1$. It is
    accepting if $c_0 = (q_0, r, \gamma_0,\epsilon)$, where $r$ is the
    root node of $t$, and $c_m =
    (q_s, n, \beta, u)$ for some node $n$ of $t$, $\beta\in\Gamma^+$,
    and $u\in\Sigma^*$. 
    Since $R$ is function,
    there exists at most one accepting run per input tree $t$, and we
    call $u$ the output of $t$. The transduction realized by $T$ is 
    the set of pairs $(t,u)$ such that there exists an accepting run of $T$ on
    $t$ whose output is $u$.

\paragraph{Equivalence between \dtwovpt and DPT$_u$WT} Modulo
nested word linearisation of unranked trees, the two models are equivalent. Let
us briefly sketch why.

Assume $T$ is a DPT$_u$WT and let us construct an
equivalent $\dtwovpt$ $T'$. First notice that when $T$ is positioned at
some node $n$, its stack height is exactly the depth of node $n$ in
the tree, as well as the depth of the call and return symbols
corresponding to $n$ in the linearisation. Also note that $T$ can
always read the top symbol of the stack, while $T'$ only reads it when
it pops a symbol. This issue can be overcome by always keeping in the
state of $T'$ the top stack symbol. It remains to see how $T'$ can simulate the
moves of $T$ and its tests (root, leaf, etc.). We assume that
if $T$ is positioned at some tree node $n$, then $T'$ is positioned at
the call position $c_n$ corresponding to $n$ in the linearisation of the
input tree. Then, if $T$ moves from $n$ to its next-sibling $n'$, $T'$ 
has to traverse the whole linearisation of the subtree rooted at
$n$. It can be easily done by pushing a special symbol when reading 
$c_n$ forward, which is popped once the matching return position of
$c_n$ is met. Simulating previous-sibling moves is done
symmetrically. Suppose now that a tree node $n'$ is the parent of a
tree node $n$. In the linearisation, it means that there is a (sub) nested
word of the form $c_{n'} w_1 c_n w_2 r_n w_3 r_{n'}$, where
$w_1,w_2,w_3$ are nested words. To simulate a move of $T'$ from $n$
to $n'$, $T'$ has to move backward from $c_n$ to $c_{n'}$, traversing
$w_1$. Again, by using a special stack symbol when traversing $w_1$,
$T'$ can detect when it reads $c_{n'}$: It is the first time it does
not popped the special stack symbol. To simulate a stay move, $T'$
just move one-step forward and one-step backward. 

Finally, we have to show how $T'$ can simulate the tests (root, leaf, etc.).
By using a special bottom stack symbol, $T'$ can know when it
is at the root. The other tests can easily be performed by $T'$: For
instance, to detect that $T'$ is positioned at a call position that
corresponds to a first-child, it suffices to go one-step backward and
check whether the previous symbol is call.

\emph{Conversely}, let $T$ be a $\dtwovpt$ whose input are assumed to
be linearisations of unranked trees. To construct an equivalent
DPT$_u$WT $T'$, one again has to show how the moves of $T$ are
simulated by moves of $T'$.

If $T$ moves forward by reading a call symbol $c_n$, then its next position
can be either that of a call symbol $c_{n'}$ (which means that $n'$ is
the first-child of $n$), or that of return symbol $r_n$ (which means
that $n$ is a leaf). Using a test, $T$ can decide whether it is at a
leaf or not. In the first case, it uses a stay transition and in the
second case, it uses a first-child transition. 

Other cases are treated similarly: For instance, if $T$ moves forward
by reading a return symbol $r_n$, then if the next symbol is a call
symbol $c_{n'}$, it means that $n'$ is the next-sibling of $n$, and if
the next symbol is a return symbol $r_{n'}$, it means that $n'$ is a
parent of $n$. Using tests, $T$ can decide what moves to perform,
either next-sibling or parent.

\bibliographystyleappendix{plainnat}
\bibliographyappendix{papers}

\end{document}